%% file: main.tex
\documentclass[sigplan,10pt]{acmart}
\renewcommand\footnotetextcopyrightpermission[1]{}

\usepackage[english]{babel}
\usepackage{blindtext}

\usepackage{tikz}
\usepackage{amsmath}
\usepackage{xspace}
\usepackage{url}
\usepackage{xurl}
\usepackage[most]{tcolorbox}
\usepackage{listings}
\usepackage{xcolor}
\usepackage{caption}
\usepackage{pifont}
\usepackage[dvipsnames]{xcolor}

\definecolor{codegreen}{rgb}{0,0.6,0}
\definecolor{codegray}{rgb}{0.5,0.5,0.5}
\definecolor{codepurple}{rgb}{0.58,0,0.82}
\definecolor{backcolour}{rgb}{0.98,0.98,0.98}

\lstdefinestyle{mystyle}{
    backgroundcolor=\color{backcolour},   
    commentstyle=\color{codegreen},
    keywordstyle=\color{magenta},
    numberstyle=\tiny\color{codegray},
    stringstyle=\color{codepurple},
    basicstyle=\ttfamily\footnotesize,
    breakatwhitespace=false,         
    breaklines=true,                 
    captionpos=b,                    
    keepspaces=true,                 
    numbers=left,                    
    numbersep=5pt,                  
    showspaces=false,                
    showstringspaces=false,
    showtabs=false,                  
    tabsize=2,
    language=Python
}

\usepackage{balance}
\usepackage[normalem]{ulem}
\usepackage{graphicx}
\usepackage{algorithm}
\usepackage{algpseudocode}
\usepackage{amsthm}
\usepackage{amsmath}
\usepackage{multirow}
\usepackage{url}
\usepackage{subcaption}
\usepackage{enumitem}
\usepackage{makecell}
\usepackage{booktabs} 
\usepackage{soul}
\usepackage{textcomp}
\usepackage{pifont}
\usepackage{adjustbox}
\usepackage{wrapfig}

\usepackage{amssymb}
\usepackage{bbm}
\usepackage{tabularx}
\usepackage{bbding}
\usepackage{tikzsymbols}
\usepackage[normalem]{ulem}

\usepackage[most]{tcolorbox}

\usepackage{tikz}

\theoremstyle{definition}

\newenvironment{denseitemize}{
	\begin{itemize}[topsep=2pt, partopsep=0pt, leftmargin=1.5em,label=$\bullet$]
		\setlength{\itemsep}{3pt}
		\setlength{\parskip}{0pt}
		\setlength{\parsep}{0pt}
	}{\end{itemize}}

\newcommand{\PHB}[1]{\noindent\textbf{#1}}
\newcommand{\PHM}[1]{\vspace{.2em}\noindent\textbf{#1}} 

\newcommand{\SystemName}{\textsc{Crab}\xspace}

\renewcommand\footnotetextcopyrightpermission[1]{} 
\setcopyright{none}

\acmDOI{}
\acmISBN{}

\begin{document}
\settopmatter{authorsperrow=5}
\title{\SystemName: A Semantics-Aware Checkpoint/Restore Runtime for Agent Sandboxes}
\settopmatter{printfolios=true}

\author{Tianyuan Wu}
\authornote{Equal contribution.}
\affiliation{
\institution{HKUST}
\country{Hong Kong SAR, China}
}

\author{Chaokun Chang}
\authornotemark[1]
\affiliation{
\institution{HKUST}
\country{Hong Kong SAR, China}
}

\author{Lunxi Cao}
\affiliation{
\institution{HKUST}
\country{Hong Kong SAR, China}
}

\author{Wei Gao}
\affiliation{
\institution{HKUST}
\country{Hong Kong SAR, China}
}

\author{Wei Wang}
\affiliation{
\institution{HKUST}
\country{Hong Kong SAR, China}
}


\begin{abstract}
Autonomous agents act through sandboxed containers and microVMs whose state
spans filesystems, processes, and runtime artifacts. Checkpoint and restore
(C/R) of this state is needed for fault tolerance, spot execution, RL rollout
branching, and safe rollback---yet existing approaches fall into two extremes:
application-level recovery preserves chat history but misses OS-side effects,
while full per-turn checkpointing is correct but too expensive under dense
co-location.

The root cause is an \emph{agent--OS semantic gap}: agent frameworks see tool
calls but not their OS effects; the OS sees state changes but lacks turn-level
context to judge recovery relevance. This gap hides massive sparsity: over
75\% of agent turns produce no recovery-relevant state, so most checkpoints are
unnecessary.

\SystemName{} (\underline{C}heckpoint-and-\underline{R}estore for \underline{A}gent Sand\underline{B}oxes) is a transparent host-side runtime that bridges this gap without
modifying agents or C/R backends. An eBPF-based inspector classifies each
turn's OS-visible effects to decide checkpoint granularity; a coordinator aligns
checkpoints with turn boundaries and overlaps C/R with LLM wait time; and a
host-scoped engine schedules checkpoint traffic across co-located sandboxes. On
shell-intensive and code-repair workloads, \SystemName{} raises recovery
correctness from 8\% (chat-only) to 100\%, cuts checkpoint traffic by up to
87\%, and stays within 1.9\% of fault-free execution time.
\end{abstract}

\maketitle
\pagestyle{plain}

\input{contents/1_Introduction}
\input{contents/2_Background}

\input{contents/3_Challenges}

\input{contents/4_Design}

\input{contents/5_Implementation}
\input{contents/6_Evaluation}
\input{contents/7_Relatedworks}

\input{contents/8_Conclusion}

\bibliographystyle{ACM-Reference-Format}
\bibliography{reference}

\clearpage


\end{document}

%% file: contents/1_Introduction.tex
\section{Introduction}
\label{sec:intro}

AI systems are undergoing a fundamental shift, evolving from passive single-turn
chatbots to \emph{autonomous multi-step agents} that reason,
act, and iterate in real execution environments~\cite{yang2024swe,
codex_overview, claude_code_overview, openclaw_sandboxing}. These agents compile
projects, run shell commands, install dependencies, start background services,
and iteratively modify files and runtime state until a task is resolved. To
support such workloads, production platforms provide \emph{sandboxed
environments}---containers or microVMs that expose a full Linux OS with a
filesystem, shell, and process space~\cite{e2b_docs, wang2024openhands}. Agent
sandboxes have become critical infrastructure: they are the medium through which
agents act on the world. Agents are therefore \emph{OS-level actors} whose
effective state spans the conversation history \emph{and} kernel-visible
state---filesystem contents, long-lived processes, and installed packages.

This rich, accumulated state makes checkpoint/restore (C/R) an increasingly
important systems capability for sandboxed agents. Three use cases are
particularly pressing. \emph{Fault tolerance:} long-horizon agent workloads can
span hundreds of turns ~\cite{agentcgroup,rollart,dmsubgoalagent,anthropic_multi_agent,openai_codex_long_horizon}; a crash without a checkpoint forces costly
re-execution of all prior work. \emph{Spot/preemptible execution:} cloud spot
instances offer substantial cost savings for sandbox deployment but can be preempted
on short notice~\cite{aws_spot_notice}; C/R preserves progress across
preemptions. \emph{RL rollouts:} tree-based reinforcement learning (RL)
post-training algorithms such as Tree GRPO~\cite{ji2025tree} branch from
intermediate states; C/R can fork a sandbox from a checkpoint instead of
re-executing a shared prefix~\cite{wang2025let}. C/R is also critical to \emph{safe rollback},
where operators restore the sandbox to a known-good state after a catastrophic
or adversarially induced action~\cite{openclaw_sandboxing, taming_openclaw}.
All of these cases require the runtime to efficiently snapshot and restore
the \emph{full sandbox state} on which agent execution depends.

However, efficient C/R for agent sandboxes is not a straightforward application
of existing mechanisms. The core challenge is an \emph{agent--OS semantic gap}:
agent execution interleaves LLM reasoning with tool calls whose effects
materialize as OS state changes, yet neither the agent layer nor the OS layer
alone has the information needed for efficient C/R. Application-level methods,
such as coding-assistant rewind~\cite{claude_code_checkpointing,
codex_overview} and framework persistence~\cite{langgraph_persistence,
autogen_state, llamaindex_state}, preserve conversational or file state cheaply
but miss OS-side effects such as installed packages, spawned processes, and
files modified by shell commands. OS- and VM-level
mechanisms~\cite{criu_main, runc_criu, firecracker-snapshot, e2b-snapshots}
guarantee correctness by saving full execution state, but they are
\emph{semantics-oblivious}: they treat every step as equally stateful, imposing
substantial cost and poor scalability under high-density sandbox deployment.
Table~\ref{tab:intro-compare} summarizes this divide; we quantify it on
realistic agent workloads in \S\ref{sec:gaps}.

\begin{table}[t]
\centering
\footnotesize
\begin{tabular}{lcccc}
\toprule
\textbf{System} & \textbf{Layer} & \textbf{State Captured} & \textbf{Correct} & \textbf{Cost} \\
\midrule
Claude~\cite{claude_code_checkpointing} & App & Chat + Git/FS & $\times$ & Low\\ 
LangGraph~\cite{langgraph_persistence} & Framework & Conversation & $\times$ & Low\\ 
\midrule
Docker~\cite{chen2015checkpoint} & OS & Full Container & $\checkmark$ & Med \\ 
E2B~\cite{e2b-snapshots} & VM & Full VM & $\checkmark$ & High \\ 
\midrule
\textbf{\SystemName{}} & Cross-layer & Adaptive & $\checkmark$ & Low \\ 
\bottomrule
\end{tabular}
\caption{Comparison of C/R approaches for agent sandboxes. App-level methods are efficient but incomplete; OS/VM-level methods are correct but costly. \SystemName{} bridges this gap with semantics-aware C/R.}
\vspace{-1cm}
\label{tab:intro-compare}
\end{table}

Bridging this gap requires a specialized C/R system for agent sandboxes that
satisfies three requirements. \textbf{R1: Recovery correctness.} Each restore point must
reconstruct the sandbox state on which future execution depends---including
filesystem and process effects of prior tool calls---not merely chat history.
\textbf{R2: Low exposed overhead.} Checkpointing must not dominate end-to-end
agent latency; a system that pauses the sandbox for every checkpoint negates the
benefit of C/R. \textbf{R3: Scalable under dense co-location.} Checkpoint traffic from
many sandboxes must not collapse shared host resources such as storage bandwidth
and C/R backends. Meeting all three requires that the system \emph{infer} which
turns produce \emph{recovery-relevant state} and \emph{coordinate} checkpoint
work across co-located sandboxes.

We present \SystemName{}, a transparent host-side runtime that bridges this gap
without modifying either the agent or existing C/R backends. Our key insight is
that the runtime can treat the agent as a black box and instead observe
\emph{OS-visible effects at turn boundaries} to decide checkpoint granularity.
Because agents naturally alternate between local tool execution and waiting for
the next LLM response, the system can overlap checkpoint work with LLM wait
time, hiding most of the cost. \SystemName{} is built from three components.
A user-space \emph{Coordinator} sits on the agent--LLM control path as a proxy, identifies
turn boundaries~(\textbf{R1}), and overlaps checkpoint work with LLM wait
windows~(\textbf{R2}). A lightweight eBPF-based~\cite{linux_ebpf_syscall} \emph{Inspector} observes
OS-visible effects after each turn---filesystem changes, process state, memory
modifications---and classifies each turn as requiring no checkpoint, a
filesystem-only checkpoint, a process-only checkpoint, or a full checkpoint~(\textbf{R1}). A host-side
\emph{C/R Engine} schedules and executes checkpoint requests across densely co-located
sandboxes using commodity backends---\texttt{runc} for lifecycle management, ZFS
snapshots~\cite{zfs_snapshots} for filesystem state, and CRIU~\cite{runc_criu}
for process state---smoothing checkpoint bursts and prioritizing jobs that would
otherwise block agent progress~(\textbf{R3}).

We evaluate \SystemName{} on
code-repair SWE-Bench~\cite{yang2024swe} and shell-intensive Terminal-Bench~\cite{terminalbench} workloads driven by three agents and
LLMs. \SystemName{} achieves 100\% recovery correctness, compared with
8--13\% for chat-only and 28--42\% for chat+filesystem baselines on
Terminal-Bench. Despite one crash per task, \SystemName{} stays within 1.9\% of
no-fault execution time---whereas restarting from scratch adds up to
1.67$\times$ on SWE-Bench, and every-turn full checkpointing adds up to
3.78$\times$ on Terminal-Bench at 96-sandbox density due to host I/O contention.
\SystemName{} classifies up to 87\% of turns as requiring no checkpoint,
eliminating corresponding C/R works entirely. In runs without injected
failures, asynchronous overlap keeps the p95 exposed checkpoint delay to 0.44\%
of task time at a density of 64 co-located sandboxes. Beyond failure recovery, \SystemName{} also \emph{enhances} the agents by enabling agent-facing uses of C/R. In our case studies, exposing rollback as a tool lets agents replace brittle shell-level self-recovery with a single sandbox restore, reducing wall-clock time by up to 29\% and rollback tokens by 36\%.

This paper makes three contributions. 
\begin{denseitemize}
    \item We identify the agent--OS semantic gap and quantify its consequences:
    chat-only recovery succeeds on only 8--13\% of Terminal-Bench tasks, while
    every-turn full checkpointing slows execution by up to 3.78$\times$ under
    dense co-location~(\S\ref{sec:gaps}).
    \item We design and implement \SystemName{}, a host-side runtime that infers
    recovery-relevant state from OS-visible effects, overlaps checkpoint work
    with LLM wait time, and coordinates C/R across co-located
    sandboxes---without modifying agents or C/R
    backends~(\S\ref{sec:design}--\S\ref{sec:runtime}).
    \item We evaluate \SystemName{} on Terminal-Bench and SWE-Bench and show
    that it achieves 100\% recovery correctness while staying within 1.9\% of
    no-fault, checkpoint-free execution time, even at 96-sandbox density~(\S\ref{sec:eval}).
\end{denseitemize}



%% file: contents/2_Background.tex
\section{Background}
\label{sec:background}

\PHB{AI Agents and Sandboxes.}
AI agents are increasingly deployed to tackle software engineering and
automation tasks that require interacting with real execution environments, not
just generating text. Agents such as Claude Code~\cite{claude_code_overview} and
Codex~\cite{codex_overview} can read and edit files, run shell commands, and
coordinate multi-step workflows across tools and repositories. Their execution
follows an interleaved control loop between the large language model
(LLM) and the execution environment: the agent receives an LLM response, issues
tool actions, collects their results, feeds the observations back to the model,
and repeats. We call one such model--environment exchange an \textit{interaction
turn}.

These agents typically run in \emph{sandboxes}: isolated execution environments
such as containers or VMs that expose a real OS interface, including a
filesystem, shell, and process space. Agent and sandbox can be integrated in two
ways. In the \emph{agent-in-a-sandbox} model, the agent process itself runs
inside the sandbox. In the \emph{agent-with-a-sandbox} model, the agent remains
outside but interacts with the sandbox through its interfaces, e.g., by
executing commands or manipulating files remotely. Both patterns are common in
deployed systems, from cloud sandboxes such as E2B~\cite{e2b_docs} to
Docker-backed environments such as OpenHands~\cite{wang2024openhands}. In
either case, each interaction turn mutates not only the conversation history but
also the OS state of the sandbox.

\begin{table}[tb]
\centering
\footnotesize
\begin{tabular}{lcccc}
\toprule
\textbf{Scenario} & \textbf{Checkpoint} & \textbf{Restore} & \textbf{\#Ckpts} & \textbf{Need} \\
\midrule
FaultRecover      & Frequent & On failure    & Single   & Fast snap \\
Spot~\cite{aws_spot_notice}  & On preempt & On restart & Single & Fast migrate \\
TreeRL~\cite{ji2025tree} & Per branch & On expand & Multi & Fast fork \\
SpecAct\cite{ye2026speculative} & Frequent & On reject & Single & Fast C\&R \\
Proactive C/R & Frequent & On demand & Multi & Tool API\\
\bottomrule
\end{tabular}
\caption{Use cases for agent C/R. They differ in checkpoint timing and persistence needs but all require efficient snapshot and recovery of sandbox state.}
\vspace{-1cm}
\label{tab:motivation-scenarios}
\end{table}

\PHM{C/R for Sandboxed Agents.}
Because sandboxed agents accumulate mutable OS state, checkpoint/restore (C/R)
is a key system capability. Efficient C/R enables fault tolerance under crashes,
cost-effective execution on preemptible spot instances, and reinforcement
learning (RL) algorithms such as Tree GRPO~\cite{ji2025tree}.
As shown in Table~\ref{tab:motivation-scenarios}, these use cases impose
different requirements on the C/R substrate. In \textit{Fault Recovery},
checkpoints must be frequent enough to avoid redoing substantial work after
failures. In \textit{SpotAgent}, checkpoints must preserve progress across
preemption events---which may arrive with little
warning~\cite{aws_spot_notice}---and support fast restore for efficient migration.
In \textit{TreeRL}~\cite{ji2025tree}, checkpoints enable branching rollouts,
where many executions share a common prefix and re-executing that prefix is
wasteful. In speculative action execution~\cite{ye2026speculative}, checkpoints let the runtime fork from the current sandbox state, execute draft actions in parallel, and quickly discard or commit the fork depending on whether the oracle action agrees. Moreover, exposing rollback as an agent-facing API enables proactive recovery, allowing the agent to replace brittle shell-level cleanup with a direct restore to a known-good checkpoint.

Although these scenarios differ in checkpoint timing, restore triggers, and
persistence needs, they share a common requirement: the runtime must snapshot
and recover sandbox state efficiently. However, no existing system addresses this well,
as we show in \S\ref{sec:gaps}.

%% file: contents/3_Challenges.tex
\section{Gaps and Challenges}
\label{sec:gaps} 

Existing C/R approaches for agents fall into two extremes, each with a
fundamental limitation. Application- and framework-level methods recover
conversational or file state but miss OS-side effects, producing inconsistent
environments (\S\ref{sec:gap-lightweight}). OS- and VM-level methods capture
full sandbox state but treat every turn as equally stateful, imposing
prohibitive cost under dense deployment (\S\ref{sec:gap-fullstate}). Both gaps
stem from an \emph{agent--OS semantic gap}: neither the agent nor the OS
layer alone has the information needed for efficient C/R. We identify this gap
and show why it cannot be bridged at the tool API level
(\S\ref{sec:gap-semantic}), motivating our system design in \S\ref{sec:design}.

\subsection{Lightweight Recovery Is Often Incorrect}
\label{sec:gap-lightweight}

Application- or framework-level approaches, such as
LangGraph~\cite{langgraph_persistence} and coding assistants like Claude
Code~\cite{claude_code_overview}, preserve
conversational history or file changes but not the live runtime state that
subsequent tool invocations may depend on. Because tool execution produces
concrete OS-level effects, including filesystem modifications and in-memory
process state, resuming from conversational or filesystem state alone often
leads to an inconsistent environment.

To quantify this gap, we randomly sample 50
Terminal-Bench~\cite{terminalbench} tasks that succeed in a clean run using the
iFlow-ROME~\cite{wang2025let} model, inject exactly one failure at a random
position along each trajectory, and compare three recovery strategies against a
\textbf{No failure} baseline: (1) \textbf{Restart}, which restarts execution from the
beginning, (2) \textbf{Chat-only}, which recovers using only the chat history, and (3) \textbf{Chat+FS}, which recovers the filesystem together with the
conversational history, but without in-memory process state. To separate
recovery requirements from model stochasticity and the agent's ability to
``repair'' a broken environment, we evaluate under both deterministic trace
replay and a live LLM service.

\begin{figure}[tb]
  \centering
  \begin{subfigure}[t]{\columnwidth}
    \centering
    \includegraphics[width=\linewidth]{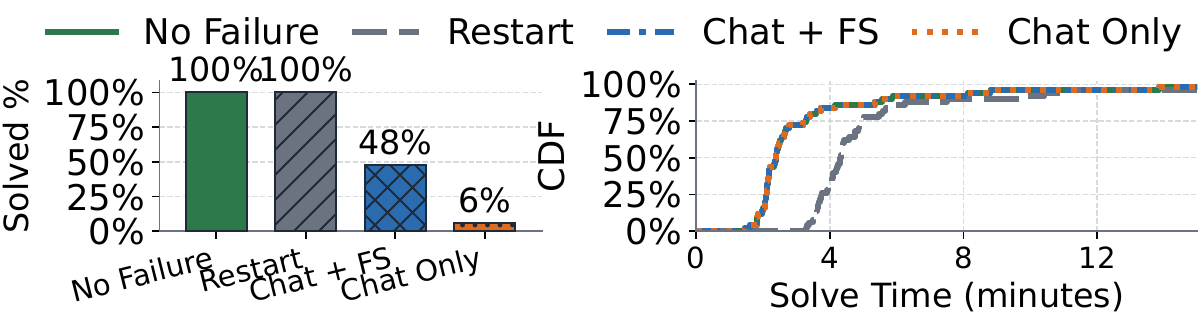}
    \caption{Replay-based evaluation.}
    \label{fig:challenge-state-replay}
  \end{subfigure}

  \vspace{2pt}

  \begin{subfigure}[t]{\columnwidth}
    \centering
    \includegraphics[width=\linewidth]{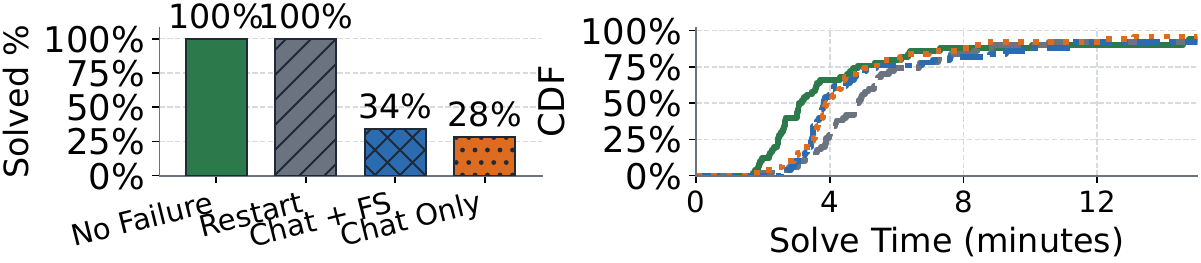}
    \caption{Real-LLM evaluation.}
    \label{fig:challenge-state-llm}
  \end{subfigure}

  \caption{Recovery-state gap under four recovery strategies. Left: task solve rate. Right: CDF of per-task runtime. Slowdown is defined as median restart / no failure solve time.}
  \Description{Visualization of recovery-state gap: left panel shows task solve rate, right panel shows CDF of per-task runtime.}
  \label{fig:challenge-state}
\end{figure}

Figure~\ref{fig:challenge-state} shows the evaluation results.
\textbf{Restart} preserves correctness but increases median time-to-solve by
1.81$\times$ (replay) and 1.55$\times$ (live LLM) because all prior work must be
re-executed. \textbf{Chat-only} succeeds on only 6\% of tasks under replay and 28\% with a
live LLM. \textbf{Chat+FS} improves replay success to 48\% but reaches only 34\% with a
live LLM: once the recovered environment diverges from the original execution,
the agent may notice missing state and take corrective actions that further
perturb the run. In short, lightweight recovery is fundamentally incomplete,
while restart-based recovery is correct only by re-executing all prior work.

\subsection{Full-State Checkpointing Does Not Scale}
\label{sec:gap-fullstate}

The opposite extreme to lightweight recovery is to checkpoint the \emph{full sandbox state} after every
turn. This approach preserves correctness but imposes prohibitive overhead. A full checkpoint
must capture both filesystem state and live process state. Even modern sandbox
substrates expose this cost: E2B~\cite{e2b_docs} (based on Firecracker~\cite{firecracker-snapshot})
requires pausing the microVM before snapshot creation and introduces more than
4\,s overhead~\cite{e2b-snapshots,firecracker-snapshot}. Even with faster
off-the-shelf Linux substrates---ZFS~\cite{zfs_snapshots} for filesystem state
and \texttt{runc}+CRIU~\cite{runc_criu} for process state---the overhead could reach seconds, dominated by process checkpointing.

\begin{figure}[tb]
  \centering
  \includegraphics[width=\linewidth]{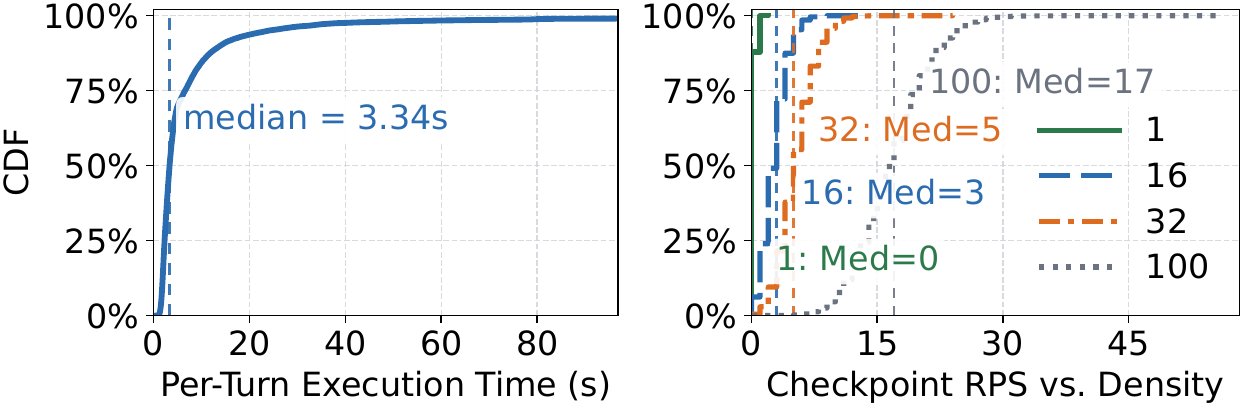}
  \vspace{-5mm}
  \caption{Agent turn-time distribution and checkpoint pressure at host scale. Left: CDF of agent turn time. Right: host-side checkpoint arrival RPS distribution vs sandbox density.}
  \Description{Agent turn-time distribution and checkpoint pressure at host scale. Left: CDF of agent turn time. Right: host-side checkpoint arrival RPS distribution vs sandbox density.}
  \label{fig:challenge-pressure}
\end{figure}

The cost compounds at host scale. Agents generate frequent checkpoint
opportunities: as Figure~\ref{fig:challenge-pressure} (left) shows, our
Terminal-Bench traces exhibit a median turn duration of 3.34\,s and 117 expected
turns per task. With 100 sandboxes per host (128 cores), naively checkpointing every turn
yields a median checkpoint arrival rate of 17 requests/s and a p90 of 26
requests/s (Figure~\ref{fig:challenge-pressure} (right)). Such bursty arrivals
introduce frequent interruptions to each sandbox and quickly overwhelm shared
C/R backends.

\begin{figure}[tb]
  \centering
  \includegraphics[width=\linewidth]{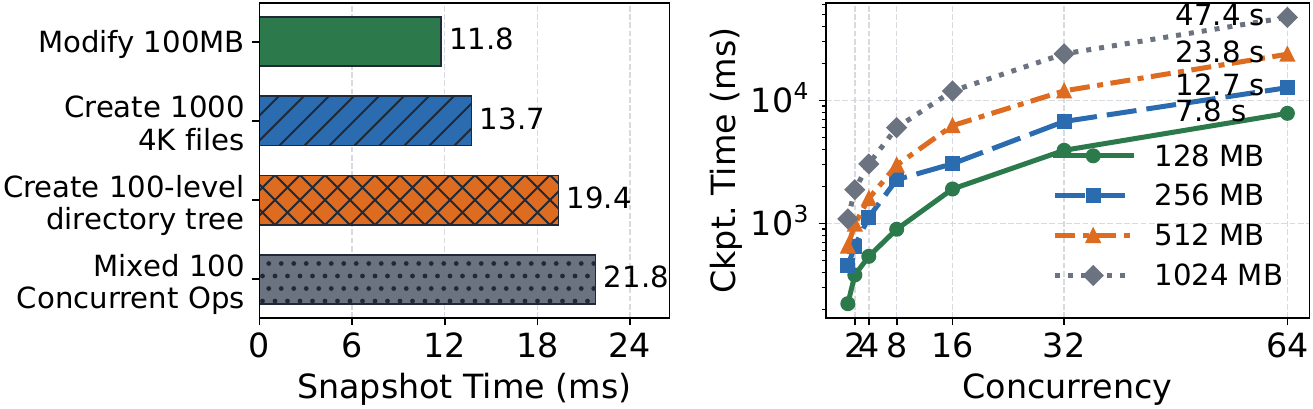}
  \caption{Left: ZFS snapshot overhead remains within tens of ms. 
  Right: CRIU-based process checkpoint latency can grow to tens of seconds under high concurrency; Testbed: AWS \texttt{c6id.32xlarge}~\cite{aws_c6id} with local NVMe SSDs.}
  \Description{Left: ZFS snapshot overhead remains within tens of ms. 
  Right: CRIU-based process checkpoint latency can grow to tens of seconds under high concurrency.}
  \label{fig:challenge-snapshot-overhead}
\end{figure}

Figure~\ref{fig:challenge-snapshot-overhead} shows the underlying bottleneck.
While ZFS snapshots~\cite{zfs_snapshots} remain cheap---even under high concurrency,
per-snapshot overhead stays within 22\,ms---process checkpoints can be expensive. 
As Figure~\ref{fig:challenge-snapshot-overhead} (right) shows,
CRIU~\cite{criu_main} dump cost grows with the live memory footprint,
which is especially problematic for agent sandboxes:
AgentCgroup~\cite{agentcgroup} reports a stable framework baseline of roughly
185\,MB, with tool-driven memory bursts that can exceed 4\,GB. In our concurrent
dump experiment on a local AWS NVMe SSD with 2.1M/1.1M random R/W IOPS, even 16 concurrent 128\,MB dumps
prolong dump time to 1.3\,s, while 64 concurrent 1\,GB dumps take 47\,s. At
realistic co-location densities (e.g., 100 sandboxes on a 128-core host), full-state, per-turn 
checkpointing easily overwhelms host I/O and cannot scale.



\subsection{The Agent--OS Semantic Gap}
\label{sec:gap-semantic}

\S\ref{sec:gap-lightweight} and \S\ref{sec:gap-fullstate} rule out both obvious
extremes: recovering only conversational or filesystem state is often incorrect,
while checkpointing full process state after every turn is too expensive. What
is missing is a way to determine \emph{whether} a turn modifies
\emph{recovery-relevant} sandbox state, and thus \emph{whether} a full checkpoint is
necessary. In real agent deployments, agent turns that make sandbox state changes are sparse, where more than 75\% of turns are stateless (e.g., file reads), see Figure~\ref{fig:checkpoint_skip_ratio} in our evaluation. 
This information, however, is not visible at either the application
or OS layer alone---a problem we call the \emph{agent--OS semantic gap}. Agent
frameworks observe tool invocations but cannot determine their concrete OS
effects. Conversely, the OS observes filesystem and process changes but lacks
the agent-level context to judge whether those changes matter for recovery.

\begin{figure}[tb]
  \centering
  \includegraphics[width=\linewidth]{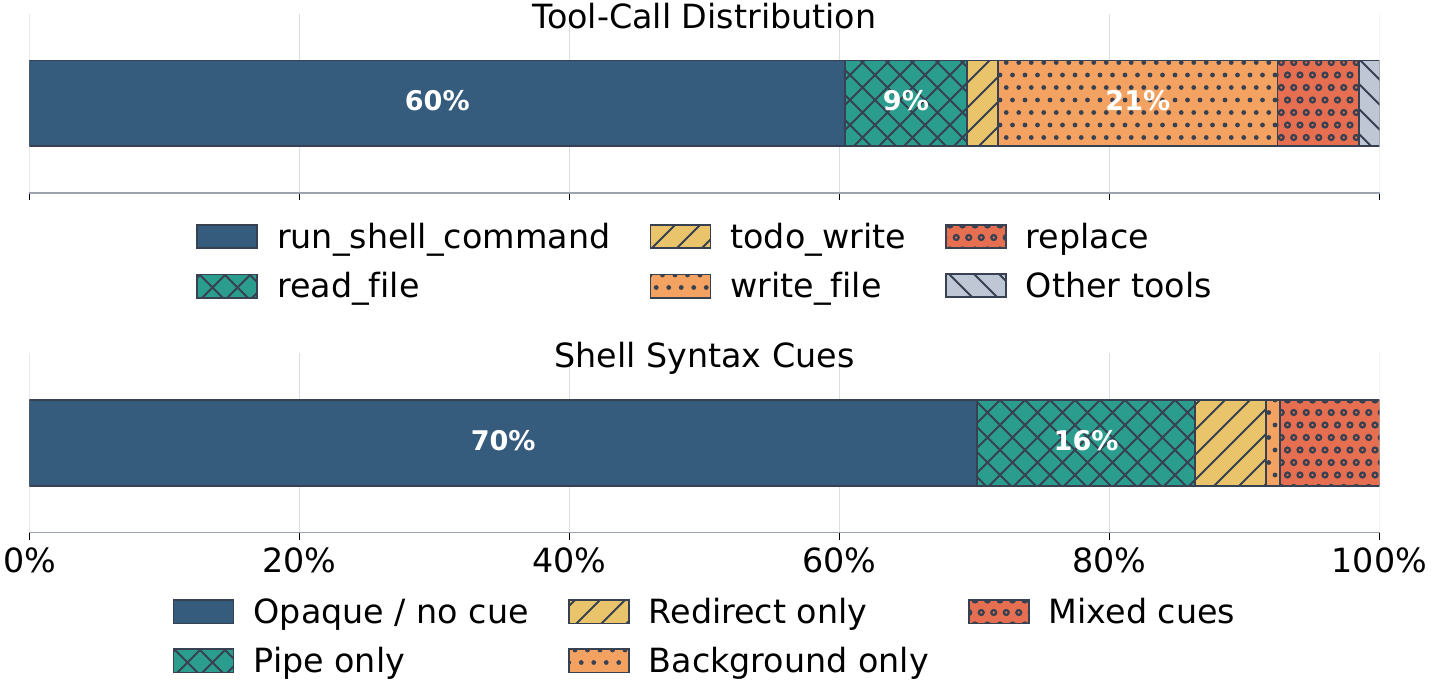}
  \vspace{-5mm}
  \caption{Opacity of agent actions. Top: tool-call distribution. Bottom: fraction of shell commands with explicit side-effect syntax versus semantically ambiguous commands.}
  \Description{Opacity of agent actions. Top: tool-call distribution. Bottom: fraction of shell commands with explicit side-effect syntax versus semantically ambiguous commands.}
  \vspace{-0.6cm}
  \label{fig:challenge-opacity}
\end{figure}

This agent--OS semantic gap cannot be bridged by inspecting tool invocations, because the dominant
tool type reveals almost nothing about its OS-level effects. In our
Terminal-Bench traces, \texttt{run\_shell\_command} accounts for 60.4\% of all
tool calls (Figure~\ref{fig:challenge-opacity}). At the API level, a call like
\texttt{python arbitrary\_script.py} may modify files, spawn long-lived
processes, or simply produce output---yet the tool signature is identical in
each case. Moreover, the command space is \emph{unbounded}: commands can
invoke arbitrary programs or scripts whose side effects cannot be reliably
inferred from syntax or static analysis. Even explicit syntactic signals are
rare: among 12,483 shell commands in our traces, only 5.3\% contain redirection,
1.0\% use background execution, and 16.2\% involve pipes. Treating all actions
as stateful recreates the cost of full-state checkpointing; relying on tool
labels or syntax alone risks the incorrect recovery shown in
\S\ref{sec:gap-lightweight}. The key challenge is therefore to infer
recovery-relevant state changes from \emph{observed OS-level effects} at turn
boundaries, rather than from tool names or command syntax---a problem we address
in \S\ref{sec:design}.

%% file: contents/4_Design.tex
\section{Design Overview}
\label{sec:design}

Bridging the agent--OS semantic gap requires a runtime that combines agent-level
turn semantics with OS-level visibility into recovery-relevant state. Such a
runtime must satisfy three requirements. \textbf{R1: Correctness.} Each
restore point must correctly reconstruct the sandbox state on which future execution
depends, including filesystem and process effects. \textbf{R2: Low exposed
overhead.} Checkpointing should be taken out of the sandbox's critical path, with minimal
impact on end-to-end execution. \textbf{R3: Scalable
under co-location.} Checkpointing must remain efficient under dense sandbox co-location,
so that contention for shared C/R backends and storage bandwidth does not
inflate task latency.

\subsection{Design Rationale}
\label{subsec:design-rationale}

These requirements motivate our semantics-aware host-side design. The key
observation is that the runtime must understand, at turn granularity, what state
has become recovery-relevant and when checkpoint work can be performed without
unnecessarily extending the critical path. Accordingly, \SystemName{} is
organized around three key ideas.

\PHM{Semantics-driven checkpointing (\textbf{R1}).}
Correct recovery requires preserving OS state that future turns depend on, which
application- or framework-level persistence alone cannot provide. At the same
time, not every turn creates recovery-relevant state, and not every stateful
turn requires the same checkpoint granularity. \SystemName{} therefore infers,
from OS-visible effects, whether a completed turn requires no checkpoint, a
filesystem checkpoint, or a full checkpoint. Capturing only the minimal
recovery-relevant state preserves correctness while avoiding unnecessary full
dumps.

\PHM{Asynchronous checkpointing at turn boundaries (\textbf{R2}).}
Even selective checkpointing can still hurt performance if its latency is
exposed on the sandbox's critical path. \SystemName{} addresses this by
exploiting the turn semantics of agent execution: agents naturally alternate
between local tool execution and waiting for the next LLM response in turns;
\SystemName{} uses this wait window to run checkpoint work asynchronously, so
much of the checkpoint cost is overlapped rather than directly added to
end-to-end latency.

\PHM{Host-level coordination of checkpoint execution (\textbf{R3}).} Checkpoint cost is paid
against shared host resources, including storage bandwidth and C/R backends.
\SystemName{} therefore schedules checkpoint work at host scope rather than
letting sandboxes checkpoint independently. This allows the runtime to smooth
bursts across co-located sandboxes and prioritize checkpoints that would
otherwise delay execution.

\subsection{Architecture and Workflow}
\label{subsec:architecture-and-workflow}

\begin{figure}[tb]
    \centering
    \includegraphics[width=\linewidth]{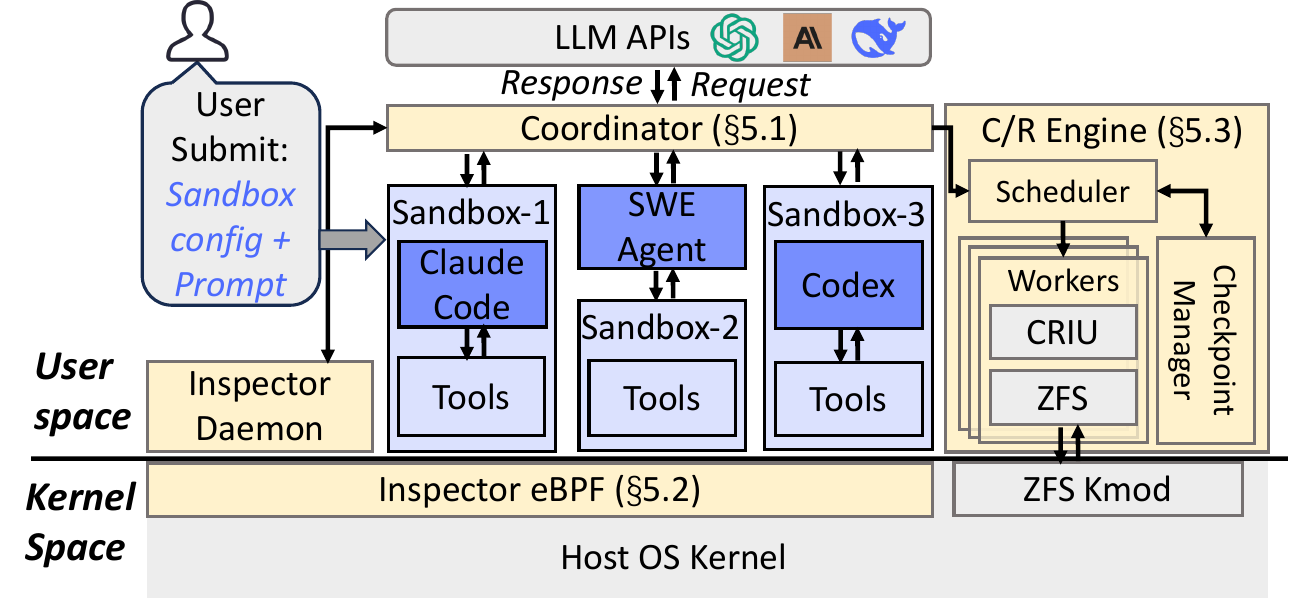}
    \caption{Architecture overview of \SystemName; yellow blocks are key components of \SystemName.}
    \Description{Architecture overview of \SystemName (yellow blocks are \SystemName's key components).}
    \label{fig:sys_overview}
\end{figure}

Realizing all three rationales requires a design point that neither a
per-sandbox ``sidecar'' nor a kernel-only approach can occupy. A per-sandbox sidecar
can react only locally and cannot coordinate checkpoint traffic across sandboxes
sharing the same host resources (\textbf{R3}). A kernel-only design can observe
OS effects but does not naturally expose turn boundaries or the LLM wait windows
needed to hide checkpoint latency (\textbf{R2}). \SystemName{} therefore adopts
a \textit{host-side runtime} that couples OS-visible observation with
host-scoped control over checkpoint timing and lifecycle.
Figure~\ref{fig:sys_overview} shows the architecture. \SystemName{} has
three key components that jointly realize all three design rationales to
meet \textbf{R1}--\textbf{R3}.

\PHM{Coordinator.}
The Coordinator sits on the control path between the agent and the
external LLM service. It identifies \textit{turn boundaries} by intercepting
requests and responses, and uses them to trigger checkpoint decisions at a
precise consistency point---realizing asynchronous checkpointing at turn
granularity (\textbf{R2}). By aligning checkpoints with completed turns, it
also establishes the consistency points required for correct recovery
(\textbf{R1}). Finally, it exposes per-sandbox execution context (e.g., whether
a checkpoint is still hidden behind the LLM wait window) to the C/R Engine,
enabling host-scoped scheduling (\textbf{R3}).

\PHM{Inspector.}
The Inspector provides the \textit{OS-level semantic signals} that
realize semantics-driven checkpointing (\textbf{R1}). After each completed
turn, it examines OS-visible effects via eBPF~\cite{linux_ebpf_syscall} and determines what
recovery-relevant state has changed: none, filesystem-only, process-only, or full state. 
This signal tells the system whether checkpointing is
needed and at what granularity, ensuring that only the minimal recovery-relevant
state is captured.

\PHM{C/R Engine.} The C/R Engine realizes host-level coordination (\textbf{R3})
and ensures that each checkpoint is a complete and consistent recovery point (\textbf{R1}).
It receives checkpoint requests from the Coordinator, schedules them across
co-located sandboxes to minimize contention and prioritize urgent work,
executes them using the underlying backends, and tracks checkpoint versions for
later restore. 


\begin{figure}[tb]
    \centering
    \includegraphics[width=\linewidth]{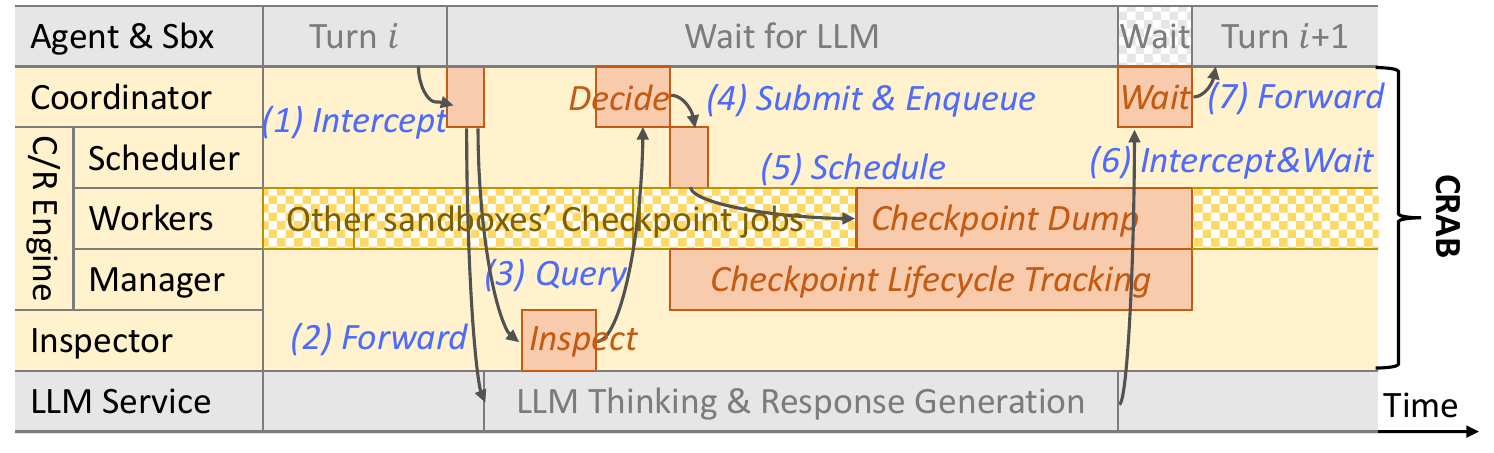}
    \caption{Workflow timeline of \SystemName.}
    \Description{Workflow timeline of \SystemName.}
    \label{fig:workflow}
\end{figure}

\PHM{Workflow.} Figure~\ref{fig:workflow} shows the steady-state
workflow, which proceeds in two phases.

\uline{\emph{1) Async checkpoint dispatch.}} At the end of turn $i$, the
\emph{Coordinator} intercepts the agent's request to the LLM (\textcircled{1})
and forwards it (\textcircled{2}). This request marks the completion of turn $i$
and opens a window in which checkpoint work can proceed without blocking the
agent. The Coordinator consults the Inspector to determine what
recovery-relevant state turn $i$ produced (\textcircled{3}) and decides whether
to \texttt{Skip} or issue a \texttt{Filesystem-only}, \texttt{Process-only}, or
\texttt{Filesystem+Process} checkpoint. If checkpointing is needed, it submits
the request for sandbox $s$ at turn $i$ to the C/R Engine
(\textcircled{4}). The C/R Engine schedules and executes this request at
host scope (\textcircled{5}): the Scheduler orders it against other
checkpoint work, Workers invoke the underlying backends, and the
Manager tracks the resulting checkpoint.

\uline{\emph{2) Completion gating.}} When the LLM response returns, the
Coordinator intercepts it before releasing it to the agent
(\textcircled{6}). If the checkpoint for turn $i$ has already completed, the
response is forwarded immediately and turn $i{+}1$ begins (\textcircled{7}).
Otherwise, the Coordinator delays release until the checkpoint finishes.
This gate ensures that no turn proceeds before its recovery state is durable,
while overlapping as much checkpoint latency as possible with LLM wait time.

%% file: contents/5_Implementation.tex
\section{Runtime and Mechanisms}
\label{sec:runtime}

We now zoom in on the three components of \SystemName{} and their runtime
mechanisms. We start to describe how the Coordinator uses an HTTP-proxy interposition to
detect turn boundaries, dispatch checkpoint work asynchronously, and gate agent
progress (\S\ref{subsec:coordinator}). We then present the Inspector's
eBPF-based state tracking and its \emph{net-change} semantics for identifying
recovery-relevant OS effects (\S\ref{subsec:inspector}). We also detail the
C/R Engine's reactive scheduling, worker dispatch, and versioned checkpoint
management (\S\ref{subsec:cr-engine}).

\subsection{Coordinator: The Control Plane}
\label{subsec:coordinator}
The Coordinator is implemented as a lightweight HTTP reverse-proxy on the
agent--LLM path. Modern agents communicate with external LLM services through
standardized HTTP APIs, so interposing at this boundary lets \SystemName{} infer
turn semantics, regulate agent progress, and enforce consistent recovery---all
without modifying agent code or checkpoint backends. Below we describe the four
mechanisms the Coordinator provides.

\PHM{Turn-Boundary Detection (R1).}
The proxy inspects every outbound HTTP request from the agent to the LLM. An
outbound request signals that the agent has finished executing the tool actions
from the previous LLM response, marking the completion of a turn. At this
point, the Coordinator records the request--response pair into a persistent
conversation log. This log serves two purposes: it defines the sequence of
turn-aligned recovery points from which the agent can resume, and it provides
the cached history needed for the fast-forward mechanism described in
\S\ref{sec:deployment}.

\PHM{Asynchronous Checkpoint Dispatch (R2).}
Immediately after forwarding the outbound request, the Coordinator queries the
Inspector (\S\ref{subsec:inspector}) for the turn's recovery-relevant state
changes and, if checkpointing is needed, submits a checkpoint job to the C/R
Engine (\S\ref{subsec:cr-engine}). Because the LLM has not yet responded, this
checkpoint work proceeds \emph{concurrently} with the LLM inference, overlapping
checkpoint latency with the LLM wait window rather than placing it on the sandbox's
critical path.

\PHM{Completion Gating (R2).} As shown in Figure~\ref{fig:workflow}
(\textcircled{6}--\textcircled{7}), the proxy buffers the returning LLM response
and checks whether the outstanding checkpoint has already finished. If so, the
response is released to the agent immediately; otherwise the proxy blocks until
the checkpoint completes. Checkpoint latency is therefore exposed to the agent
only when it exceeds the LLM wait window.

\PHM{Urgency Signaling (R3).}
If the LLM response returns while the corresponding checkpoint is still in
progress, the Coordinator promotes that job from the C/R Engine's normal queue
to its high-priority queue. This signal lets the host-side scheduler
(\S\ref{subsec:cr-engine}) prioritize checkpoint work that has become exposed on
the sandbox's critical path, while deferring jobs whose latency is still hidden
behind their LLM wait windows.

\subsection{Inspector: The State Tracker}
\label{subsec:inspector}
The Inspector realizes semantics-driven checkpointing (\textbf{R1}) by
determining, for each sandbox, whether recovery-relevant OS state---filesystem
or process---has changed since the last checkpoint. It consists of two parts: a
lightweight in-kernel eBPF~\cite{linux_ebpf_syscall} monitor that captures
OS-visible effects, and a userspace daemon that exports these results through an
HTTP query interface (Figure~\ref{fig:sys_overview}). The Coordinator queries this interface at the end of each
turn (\S\ref{subsec:coordinator}) to decide whether to skip the checkpoint,
issue a filesystem/process-only checkpoint, or request a full checkpoint. The Inspector
does not decide \emph{when} to checkpoint; it reports \emph{what} has changed.

\PHM{Net-change Semantics.}
Not every OS event that the Inspector observes is recovery-relevant. Agent
execution produces many \emph{transient} effects: Linux shells execute all
commands by forking short-lived subprocesses, and some tool invocations create
temporary files that are removed within the same turn. If every such event were
treated as a state change, nearly every turn would appear changed, even without
lasting effects.

\begin{figure}[tb]
    \centering
    \includegraphics[width=0.85\linewidth]{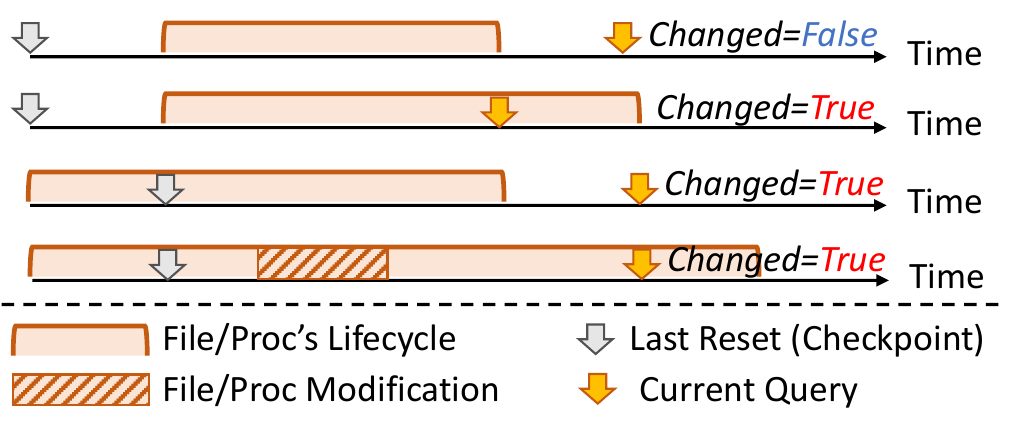}
    \vspace{-3mm}
    \caption{Examples of net filesystem / process changes.}
    \Description{Examples of net filesystem / process changes.}
    \label{fig:net_change_semantics}
\end{figure}

The Inspector therefore defines ``\emph{change}'' as the \emph{net change} between
the last checkpoint and the current inspection point
(Figure~\ref{fig:net_change_semantics}). Transient effects that have already
disappeared are ignored; only persistent effects---a file that remains created
or modified, a process that remains alive, or memory pages dirtied by a
long-running process---are reported. This semantics focuses the Inspector on
exactly the state that must be preserved for correct recovery. We implement it
with two complementary tracking mechanisms.

\PHM{Tracking Filesystem State.} We use eBPF to build a lightweight
in-kernel monitor attached to the syscall raw tracepoints
(\texttt{sys\_enter}/\texttt{sys\_exit}), which provide per-syscall argument
visibility at low overhead. The monitor operates at \textit{per-file
granularity} and records filesystem-affecting operations issued by processes in
the sandbox, including file creation, deletion, renaming, and writes. Operations
to special file handles such as \texttt{stdout} are excluded. For each event,
the monitor stores the syscall type and relevant arguments in BPF maps, which
are periodically consumed by a userspace daemon on the host. The daemon computes
the \emph{net} filesystem change since the last checkpoint. Once a checkpoint
completes, the daemon clears the accumulated BPF-map entries, resetting the
filesystem tracking baseline.

\PHM{Tracking Process and In-Memory State.} We track process-space
changes using Linux \texttt{cgroups}~\cite{linux_cgroups} together with the kernel's
soft-dirty page tracking mechanism~\cite{linux_soft_dirty}. Each sandbox is
associated with a \texttt{cgroup}, which defines the membership boundary for the
processes belonging to that sandbox. Between checkpoints, the Inspector records
whether the \texttt{cgroup} has gained new processes or lost existing ones, capturing
process creation and exit as persistent state changes.

For processes that remain alive across the interval, we detect memory
modification via soft-dirty tracking. The Inspector reads each living process's
dirty bits in \texttt{/proc/PID/pagemap} to determine whether the process has
written to any virtual memory page since the last checkpoint. Once a checkpoint
completes, the Inspector clears the soft-dirty bits via
\texttt{/proc/PID/clear\_refs}, resetting the memory
tracking baseline symmetrically with the filesystem reset described above.

\subsection{C/R Engine: The Data Plane}
\label{subsec:cr-engine}

The C/R Engine schedules checkpoint execution under dense
co-location (\textbf{R3}) and ensures that only complete, consistent recovery
points are published (\textbf{R1}). As Figure~\ref{fig:sys_overview} shows, it
comprises three sub-components: the \emph{Scheduler} orders checkpoint jobs
across co-located sandboxes to minimize exposed latency; the \emph{Workers}
execute checkpoint and restore operations using commodity backends; and the
\emph{Manager} assembles partial artifacts into versioned, recoverable
manifests.

\PHM{Scheduler.}
Under dense co-location, checkpoint traffic from many sandboxes can arrive in
bursts that exceed the available LLM wait windows
(\S\ref{subsec:design-rationale}). Because LLM wait windows are not predictable
in advance, \SystemName{} uses a \emph{reactive} scheduling policy based on
whether each job's checkpoint latency is still hidden or has already become
exposed.

The Scheduler maintains two FIFO queues: a \emph{normal} queue for checkpoint
jobs still covered by an outstanding LLM wait window, and a
\emph{high-priority} queue for jobs whose matching LLM response has already
returned. New jobs enter the normal queue, since their latency can still be
overlapped. If the LLM response returns before the checkpoint finishes, the
Coordinator promotes that job to the high-priority queue
(\S\ref{subsec:coordinator}). Workers always prefer the high-priority queue.
Starvation cannot occur: every pending job is eventually promoted when its LLM
response arrives, or has already completed in the normal queue before promotion
is needed. This policy directs limited host I/O bandwidth to jobs whose delay is
already visible to the agent, while deferring jobs whose latency can still be
hidden.


\PHM{Workers.} Workers are concurrent processes that perform the actual
checkpoint and restore operations using off-the-shelf C/R backends:
CRIU~\cite{criu_main} for process state and OpenZFS~\cite{zfs_snapshots} for
filesystem state. The number of Workers is sized to saturate the host's I/O
bandwidth without overwhelming it, maximizing checkpoint throughput under
concurrency. Each Worker receives a job from the Scheduler and executes it
according to its checkpoint type. Filesystem-only checkpoints invoke a ZFS
snapshot, which provides fast incremental checkpointing via block-level
copy-on-write. Process checkpoints invoke CRIU dump through \texttt{runc}, with
incremental checkpointing enabled when applicable. Full checkpoints compose the
two. For restore, Workers invoke the corresponding ZFS rollback and CRIU load
operations to reconstruct the requested sandbox state.

\PHM{Manager.} Because \SystemName{} may checkpoint only filesystem state,
only process state, or both, individual dump artifacts are often not valid
recovery points on their own. The Manager therefore treats recovery state as a
\emph{versioned manifest} rather than a single artifact. Each published
checkpoint version is a tuple $C_i=(P_j, F_k)$, where $P_j$ and $F_k$ denote
the most recent process and filesystem artifacts that together form a
recoverable sandbox state.

\begin{figure}[tb]
    \centering
    \includegraphics[width=\linewidth]{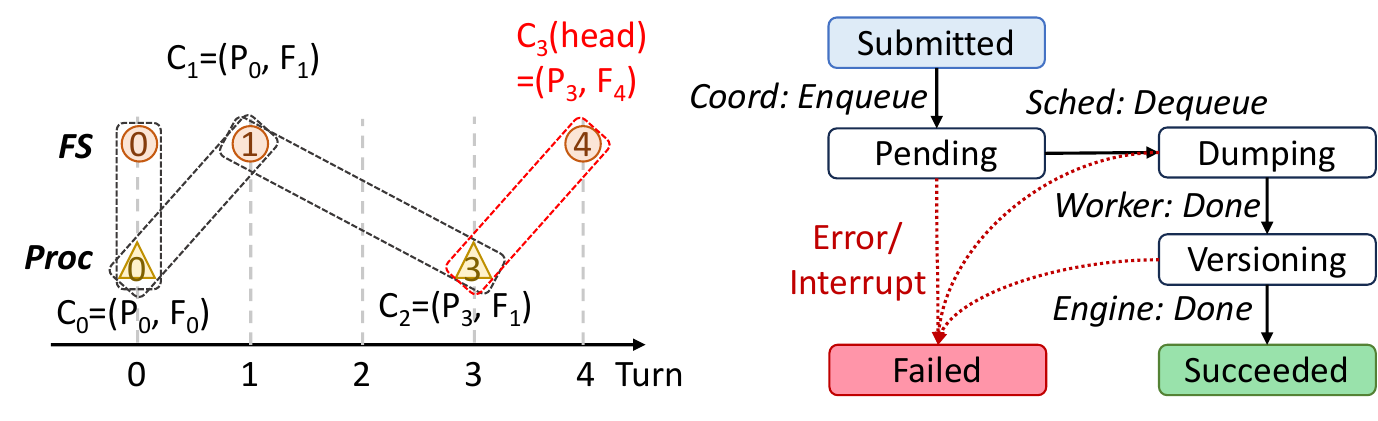}
    \caption{The Checkpoint Manager maintains versioned recoverable manifests over partial filesystem and process checkpoints (\textbf{left}), and enforces transactional checkpoint publication through lifecycle tracking (\textbf{right}).}
    \Description{The Checkpoint Manager maintains versioned recoverable manifests over partial filesystem and process checkpoints (\textbf{left}), and enforces transactional checkpoint publication through lifecycle tracking (\textbf{right}).}
    \label{fig:checkpoint_manager}
\end{figure}

Figure~\ref{fig:checkpoint_manager} (left) illustrates this versioning. When a
turn checkpoints both process and filesystem state, the Manager publishes a
full manifest (e.g., $C_0=(P_0,F_0)$). When a subsequent turn modifies only one
component, the new artifact is paired with the latest valid counterpart (e.g.,
$C_1=(P_0,F_1)$); turns that produce no recovery-relevant change leave the
manifest unchanged. In this way, the Manager maintains a git-like version
history of recoverable states even though individual checkpoint operations
update only part of the sandbox state. This abstraction supports rollback to any
earlier version, readily enabling failure recovery, spot migration, and
tree-based RL (Table~\ref{tab:motivation-scenarios}).

The Manager also ensures that checkpoint publication is \emph{transactional}
(\textbf{R1}). As shown in Figure~\ref{fig:checkpoint_manager} (right), a
checkpoint request progresses through a lifecycle: ``\emph{pending}'' upon
arrival, ``\emph{dumping}'' once assigned to a Worker, then
``\emph{versioning}'' where the Manager combines the new artifact with the
latest compatible counterpart and constructs a new manifest. Only after
versioning completes is the checkpoint marked ``\emph{done}'' and published as
recoverable. If interruption or failure occurs at any point, the checkpoint is
marked ``\emph{failed}'' and never exposed as a recovery point. This prevents
partially written or mismatched artifacts from being used during restore.

\section{Deployment Refinement}
\label{sec:deployment}

The core runtime described above supports both deployment models introduced in
\S\ref{sec:background}: \emph{agent-with-a-sandbox} and
\emph{agent-in-a-sandbox}. Each model, however, raises a distinct recovery
challenge. In the agent-with-a-sandbox model, a sandbox crash can leave an
in-flight command unfinished, requiring the runtime to mask the failure from the
external agent. In the agent-in-a-sandbox model, the agent's own in-memory state
may lag behind the restored filesystem, requiring the runtime to reconcile the
two. This section describes the mechanisms that address each challenge to ensure
correct, transparent recovery (\textbf{R1}).

\PHM{Agent-with-a-Sandbox.} In the
\emph{agent-with-a-sandbox} model, the agent runs outside the sandbox and
interacts with it through remote execution interfaces such as terminal or
command APIs (e.g., SWE agent in Figure~\ref{fig:sys_overview}). If the sandbox
crashes or is preempted while a command is \emph{in flight}, the agent observes a
failed or non-returning operation, whereas the restored sandbox has no record of
that unfinished command because checkpoints are taken only at turn boundaries.
Exposing this failure directly would break transparency and force the agent to
reason about sandbox crashes.

\SystemName{} hides this mismatch by presenting a \textit{reliable execution
interface}. Before forwarding each sandbox command, the Coordinator logs it in
its persistent conversation record. If the sandbox is restored before the
command completes, the Coordinator detects the outstanding entry and reissues
the command against the recovered sandbox. The agent therefore remains transparent 
to the failure: from its perspective, the issued command simply returns its
intended result, preserving transparent recovery (\textbf{R1}).

\PHM{Agent-in-a-Sandbox.} In the
\emph{agent-in-a-sandbox} model, the agent itself is a long-lived process inside
the sandbox (e.g., Claude Code and Codex agents in
Figure~\ref{fig:sys_overview}). If the Inspector were to track the agent as an
ordinary sandbox process, nearly every turn would appear to modify process
memory, forcing an expensive process checkpoint at turn frequency and defeating selective
checkpointing.

\begin{figure}[tb]
    \centering
    \includegraphics[width=0.9\linewidth]{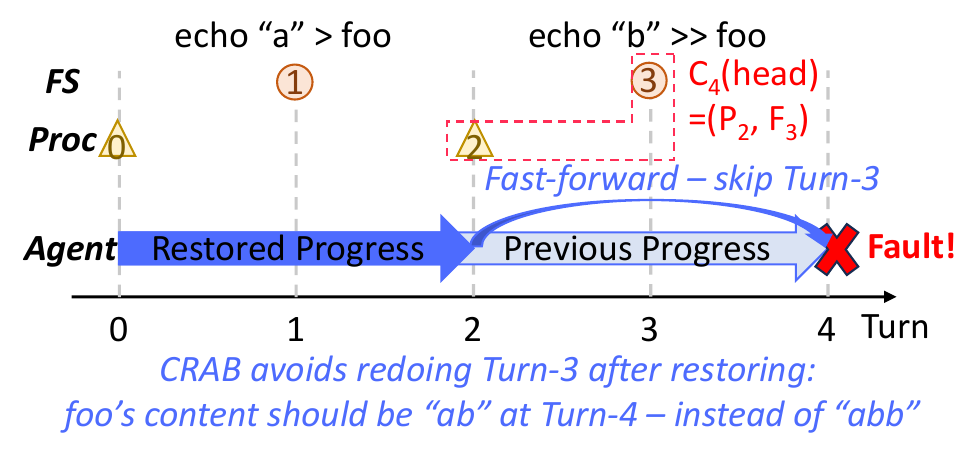}
    \caption{An example of fast-forwarding to handle the agent-in-a-sandbox recovery, where Turn-3 must be skipped as it otherwise makes the sandbox states inconsistent.}
    \Description{An example of fast-forwarding to handle the agent-in-a-sandbox recovery, where Turn-3 must be skipped as it otherwise makes the sandbox states inconsistent.}
    \label{fig:restore_in_sandbox}
\end{figure}

\SystemName{} therefore excludes the agent process from process-state tracking.
This avoids unnecessary and expensive process checkpoints but creates a
\emph{consistency problem} during restore: the checkpoint manifest may pair a
newer filesystem state with an older agent in-memory state. For example, if the
latest manifest combines process state from turn~2 with filesystem state from
turn~3 (Figure~\ref{fig:restore_in_sandbox}), the restored agent resumes at
turn-2 progress while observing a turn-3 filesystem. Naively continuing would
replay already-completed actions and corrupt the restored state.

\SystemName{} resolves this with a \emph{fast-forward mechanism}
(\textbf{R1}). The Coordinator caches all prior request--response pairs. After
restore, when the stale agent replays an earlier request, the Coordinator
recognizes it as a repeated interaction and returns a synthetic response from the
cached history instead of forwarding it to the LLM. This process repeats until
the agent's logical progress catches up with the checkpoint head
(Figure~\ref{fig:restore_in_sandbox}), restoring a consistent execution state
without re-executing any actual actions.

%% file: contents/6_Evaluation.tex
\section{Evaluation}
\label{sec:eval}

We evaluate whether \SystemName{} can provide correct, low-overhead
checkpoint/restore for densely co-located agent sandboxes. Our experiments are
driven by four questions. \textbf{Q1 (correctness and end-to-end overhead):}
Does \SystemName{}'s selective checkpointing preserve full recovery correctness,
and what end-to-end overhead does it add relative to fault-free execution?
(\S\ref{subsec:eval-correctness}) \textbf{Q2 (component overhead):} What
per-turn latency do the Coordinator, Inspector, and checkpoint execution each
contribute? (\S\ref{subsec:eval-overhead}) \textbf{Q3 (mechanism effectiveness):} How effectively does asynchronous checkpointing hide checkpoint
cost behind LLM wait time, and how much does the reactive scheduler help under
contention? (\S\ref{subsec:eval-async}) 
\textbf{Q4 (case studies):} Can \SystemName{} also \emph{enhance} agentic execution and enable proactive rollback, spot execution, speculative execution, and RL rollout branching? (\S\ref{subsec:eval-case-study})



\subsection{Experimental Setup}
\label{subsec:eval-setup}

\PHB{Host Platform.} We evaluate \SystemName{} on an Amazon AWS
\texttt{c6id.32xlarge}~\cite{aws_c6id} instance with 128 Intel Xeon Platinum 8375C cores, 256\,GB of memory, and $4\times1.7$\,TB NVMe SSDs (2.1M/1.1M Rand R/W IOPS).
\SystemName{} uses commodity C/R backends: runc-CRIU (v1.3.4), and
OpenZFS (v2.4.1).


\PHM{Agent Configurations.} Table~\ref{tab:eval_setup} summarizes our three
evaluation configurations, spanning both \emph{agent-in-a-sandbox} (Claude-code~\cite{claude_code_overview},
iFlow-cli~\cite{iflowcli} on Terminal-Bench~\cite{terminalbench}) and \emph{agent-with-a-sandbox} (SWE-agent on
SWE-bench~\cite{yang2024swe}) deployment settings. We treat all agents as black boxes and do not
modify their internal logic.

\begin{table}[tb]
    \centering
    \footnotesize
    \begin{tabular}{cccc}
        \toprule
        \multicolumn{2}{c}{\textbf{Agent}} & \multirow{2}{*}{\textbf{Benchmark}} & \multirow{2}{*}{\textbf{LLM}} \\
        \cline{1-2}
        \textbf{Name} & \textbf{Deployment} & & \\
        \midrule
        Claude-code & in sandbox & Terminal-Bench & Claude4.6-Opus~\cite{claude46} \\
        iFlow-cli & in sandbox & Terminal-Bench & iFlow-ROME~\cite{wang2025let} \\
        SWE-agent & w/ sandbox & SWE-Bench & MiniMax-M2.7~\cite{minimax27} \\
        \bottomrule
    \end{tabular}
    \caption{Agents, benchmarks, and LLMs used in evaluation.}
    \label{tab:eval_setup}
\end{table}

\begin{figure}[tb]
\centering
\includegraphics[width=0.9\linewidth]{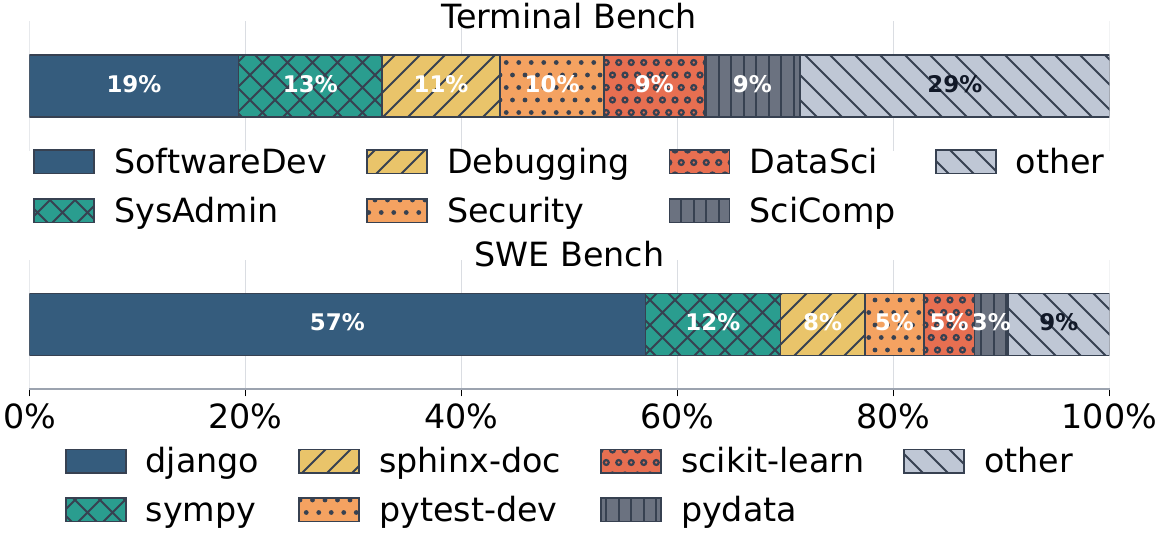}
\caption{Task-category composition of the sampled SWE-Bench and Terminal-Bench workloads.}
\Description{Task-category composition of the sampled SWE-Bench and Terminal-Bench workloads.}
\label{fig:task_category_composition}
\end{figure}

\PHM{Benchmark Characteristics.} The two benchmarks used in our study differ in what state matters
for correctness, which directly determines how much C/R must preserve.
Terminal-Bench~\cite{terminalbench} covers diverse terminal-centric tasks
(Figure~\ref{fig:task_category_composition}) whose success criterion depends on
the \emph{full sandbox state}: tasks are validated inside the same sandbox using
unit tests and interactive checks, so intermediate files, processes, and runtime
state all matter. SWE-bench~\cite{yang2024swe} focuses on resolving GitHub
issues in real software projects. Its success criterion depends primarily on the
\emph{final patch}, making intermediate runtime state less important.

\begin{figure}[tb]
\centering
\includegraphics[width=\linewidth]{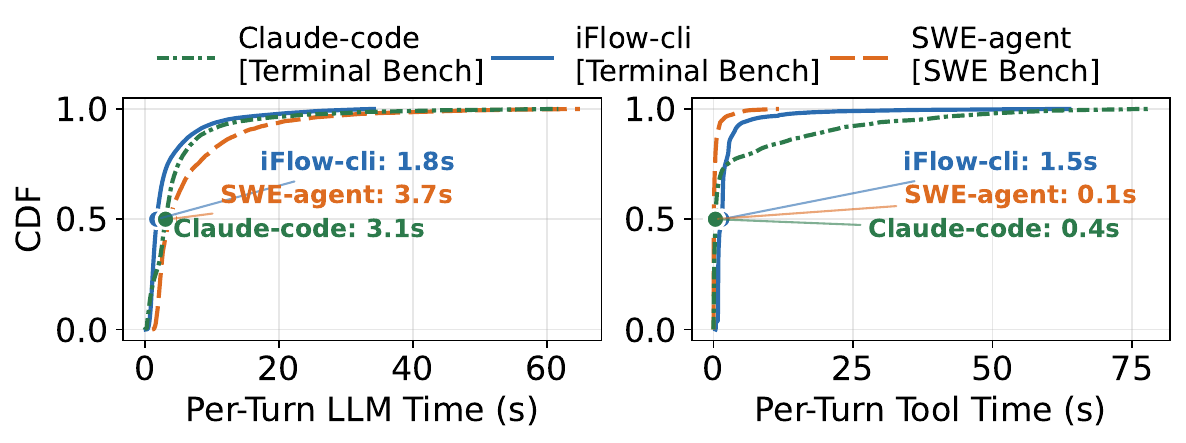}
\vspace{-7mm}
\caption{LLM and tool execution duration distributions.}
\Description{LLM and tool execution duration distributions.}
\label{fig:llm_tool_distribution_cdf}
\end{figure}

\PHM{Methodology.} For each benchmark, we randomly
sample 100 tasks that the corresponding agent \textit{can solve} in a
failure-free run. This filter isolates system-level C/R overhead from agent
capability: unsolvable tasks would introduce failures unrelated to
checkpointing and add noise to the measurements. We then record the full
agent--LLM interaction trace for each task, including model outputs, reasoning
traces, and tool invocations. Our experiments \textit{replay}
these recorded traces \emph{deterministically}. This removes LLM stochasticity and
isolates the effect of checkpointing, restore, and scheduling decisions from
variation in model behavior. The trace details are shown in
\autoref{fig:llm_tool_distribution_cdf}, where Terminal-Bench is tool-heavy due
to its complex shell interactions and SWE-bench is LLM-heavy with only
lightweight tools used.

\PHM{Baselines.} We compare \SystemName{} against four baselines that span a
range of recovery fidelity and cost. The first two reflect recovery strategies
common in today's agent frameworks; the latter two are correct but expensive
alternatives.
\begin{itemize}[leftmargin=*, nosep]
  \item \textbf{Chat-only} persists only the chat history after each turn and
restores no filesystem or process state. This is the default recovery mode in
most hosted agent platforms.
  \item \textbf{Chat+FS} persists conversational history together with filesystem
state, but not process or in-memory state. This approximates frameworks that
snapshot the workspace directory between turns.
  \item \textbf{Restart} restarts a failed task \textit{from scratch} and
re-executes the lost trajectory after sandbox failures.
  \item \textbf{Every-turn checkpoint (FullCkpt)} performs a full
filesystem+process checkpoint at \textit{every turn boundary}.
\end{itemize}


\subsection{Correctness and End-to-End Overhead}
\label{subsec:eval-correctness}

We evaluate \SystemName{} under crash recovery, the primary target use case of
the C/R capability for agent sandboxes. Each task trajectory is injected with
one crash at a random position; all runs use the same random seed.

\begin{figure}[tb]
\centering
\includegraphics[width=\linewidth]{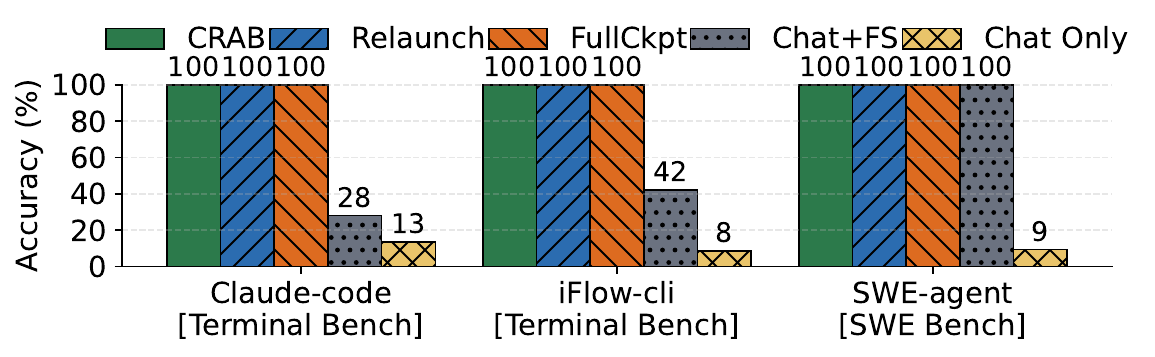}
\vspace{-7mm}
\caption{Recovery correctness under sandbox crashes.}
\Description{Recovery correctness under sandbox crashes.}
\label{fig:correctness}
\end{figure}

\PHM{Recovery Correctness.} Figure~\ref{fig:correctness}
compares \SystemName{} with the baselines. A run is counted as correct if it
satisfies the benchmark's success criterion: the entire sandbox state for
Terminal-Bench, or the final patch correctness for SWE-bench. \SystemName{},
Restart, and FullCkpt all achieve 100\% correctness in all
settings. Restart does so by restarting the task from scratch after failure;
FullCkpt preserves a recovery point at every turn boundary;
and \SystemName{} selectively checkpoints only the state that the Inspector
identifies as changed (accuracy detailed below).

In contrast, the lightweight baselines are substantially less robust. Chat+FS reaches 100\%
on SWE-bench, but only 28\% and 42\% on Terminal-Bench using Claude-code and
iFlow-cli. The key difference is that SWE-bench trajectories rarely create
long-lived processes whose state must survive recovery, so restoring the
filesystem is usually sufficient. Terminal-Bench, however, frequently
depends on background or persistent process state, making filesystem-only
recovery incomplete. Chat-only performs the worst, achieving only 13\% and 8\% on Terminal-Bench
(Claude-code and iFlow-cli) and 9\% on SWE-bench, because restoring conversation
alone treats the sandbox as stateless and loses both filesystem and process
state after a crash.

\PHM{Inspector Accuracy.}
\SystemName{} maintains full correctness because the Inspector reliably
identifies which recovery-relevant state changed at each turn.
Table~\ref{tab:inspector_accuracy} compares the Inspector against manual labels
over 2,063 turns from 100 iFlow-cli with Terminal-Bench tasks. The Inspector has
\emph{zero} false negatives (FNs) for both process and filesystem changes,
ensuring that no required state is missed. Process change detection is exact,
with 100\% accuracy. Filesystem detection reaches 98.3\% accuracy, with a 2.3\%
false positive rate (FPR) and zero false negatives. These false positives (FPs)
arise from conservative file-level tracking: if a turn writes content to a file
and later deletes it, the file is still recorded as changed. This
over-approximation is acceptable because it only causes an extra filesystem
checkpoint, which is inexpensive (only tens of milliseconds, as shown in
Figure~\ref{fig:challenge-snapshot-overhead} (left)).

\begin{table}[tb]
    \centering
    \footnotesize
    \begin{tabular}{cccccc}
        \toprule
         & \textbf{Exact} & \textbf{Detected} & \textbf{Acc.$\uparrow$} & \textbf{FPR$\downarrow$} & \textbf{FNR$\downarrow$} \\
        \midrule
        Proc. Change & 4.8\% (99) & 4.8\% (99) & 100.0\% & 0.0\% & 0.0\%\\
        FS Change & 27.9\% (575) & 29.5\% (609) & 98.3\% & 2.3\% & 0.0\% \\
        \bottomrule
    \end{tabular}
    \caption{Inspector accuracy on 2,063 manually labeled turns (iFlow-cli, Terminal-Bench). Acc=(TP+TN)/All; FPR=FP/(FP+TN); FNR=FN/(FN+TP).}
    \label{tab:inspector_accuracy}
\end{table}

\begin{figure*}[t]
  \centering
  \begin{minipage}{\linewidth}
      \centering
      \includegraphics[width=\linewidth]{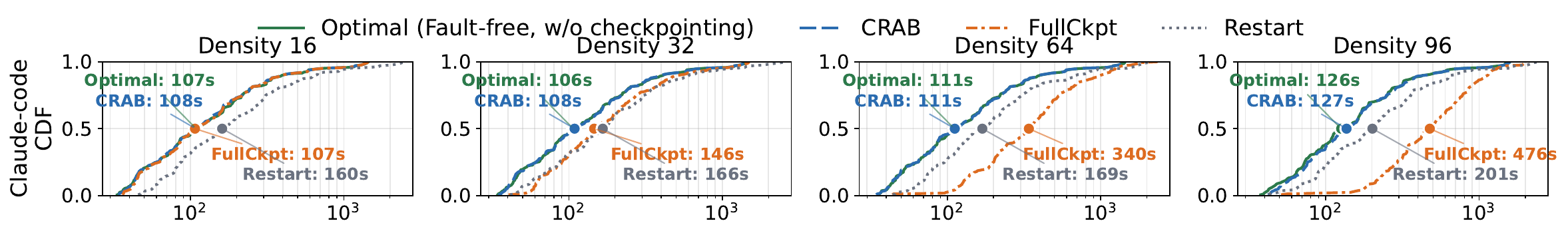}
      \vspace{-2mm}
      \label{fig:claude_code_cdf}
  \end{minipage}

  \vspace{-4mm}

  \begin{minipage}{\linewidth}
      \centering
      \includegraphics[width=\linewidth]{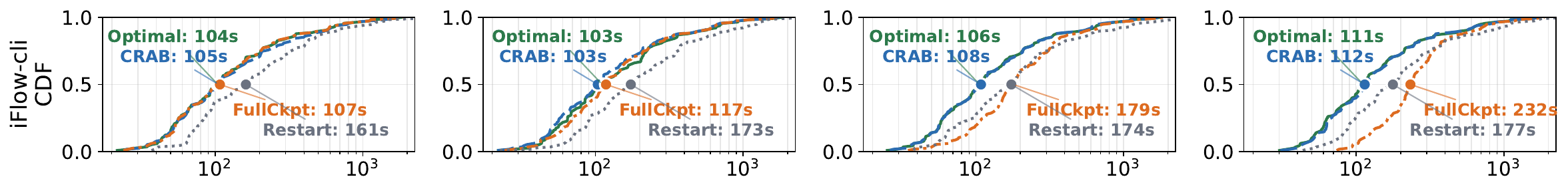}
      \vspace{-2mm}
      \label{fig:iflow_cli_cdf}
  \end{minipage}

  \vspace{-4mm}

  \begin{minipage}{\linewidth}
      \centering
      \includegraphics[width=\linewidth]{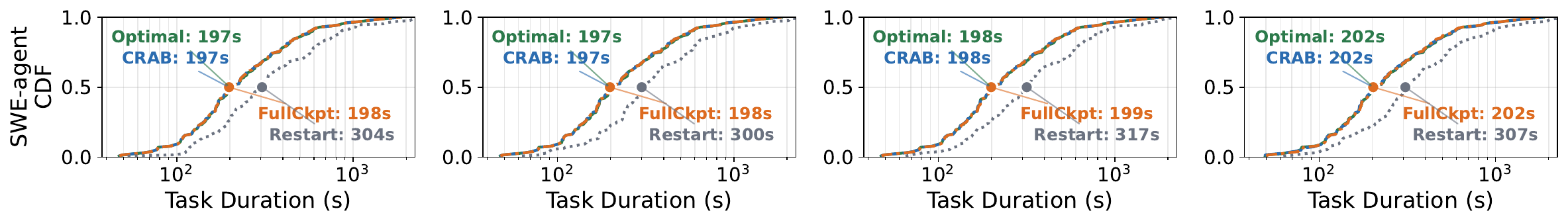}
      \vspace{-8mm}
      \label{fig:swe_agent_cdf}
  \end{minipage}
  \caption{End-to-end performance comparison across benchmarks and deployment densities under a crash recovery scenario for Claude-code, iFlow-cli, and SWE-agent. \SystemName{} remains within 1.9\% of the no-fault, checkpoint-free optimal.}
  \Description{End-to-end performance comparison across benchmarks and deployment densities under a crash recovery scenario. Top: Terminal-Bench. Bottom: SWE-bench. \SystemName{} remains within 1.9\% of the no-fault, checkpoint-free optimal.}
  \label{fig:end2endperf}
\end{figure*}

\PHM{End-to-End Overhead.} We next measure end-to-end
task completion time, from the beginning of a task's first turn to the end of
its last turn, under deployment densities from 16 to 96 co-located sandboxes per
host. We only compare \SystemName{} against baselines that ensure recovery
correctness, since it is meaningless to measure performance of incorrect
baselines. Figure~\ref{fig:end2endperf} shows that, despite one crash and
recovery per task, \SystemName{} stays within 0--1.9\% of an optimal no-fault
execution on both Terminal-Bench and SWE-bench.

Restart is substantially slower, increasing completion time by up to
1.52$\times$ on Terminal-Bench and 1.67$\times$ on SWE-bench, because a random
crash wastes roughly half of the prior work in expectation and also incurs
sandbox restart overhead. FullCkpt performs well on SWE-bench,
where turns almost exclusively modify files and filesystem checkpoints are
cheap. On Terminal-Bench, however, concurrent full checkpoints become expensive
because process state must also be preserved. For example, at densities of 64 and 96
sandboxes per host, FullCkpt for Claude-code slows execution by 3.06$\times$ and
3.78$\times$ due to host storage bandwidth contention, respectively, even worse
than restarting from scratch.

\begin{figure}[tb]
  \centering
  \includegraphics[width=\linewidth]{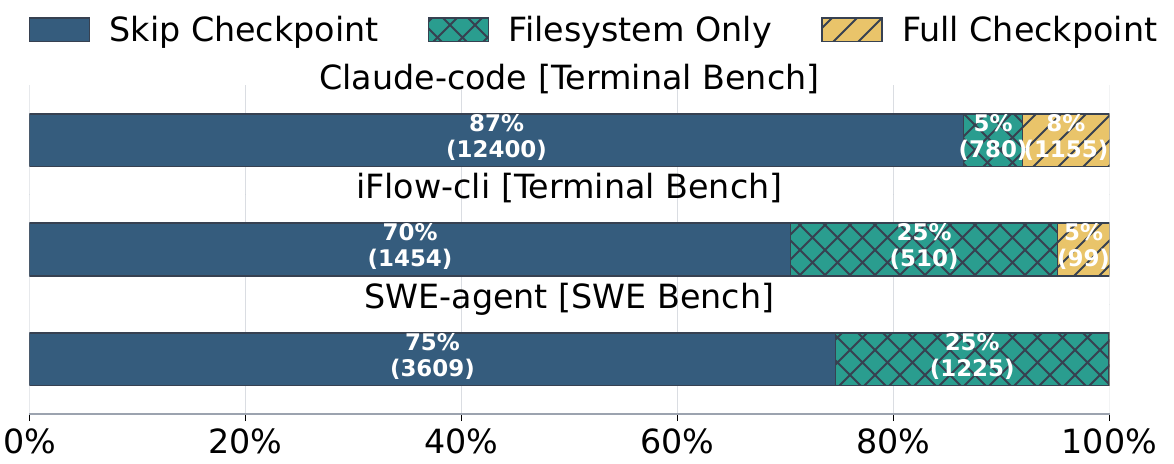}
  \vspace{-7mm}
  \caption{Checkpointing is highly sparse in \SystemName{}, with skip ratio greater than 70\%.}
  \Description{Checkpointing is highly sparse in \SystemName{}, with skip ratio greater than 70\%.}
  \label{fig:checkpoint_skip_ratio}
\end{figure}

\PHM{Checkpoint Sparsity.}
\SystemName{}'s performance advantage comes from selective checkpointing. As
shown in Figure~\ref{fig:checkpoint_skip_ratio}, up to 87\% of turns are skipped
entirely because they change neither process nor filesystem state. On
Terminal-Bench, Claude-code requires filesystem checkpoints for only 5\% of
turns and full checkpoints for 8\%; iFlow-cli requires filesystem checkpoints
for 25\% and full checkpoints for 5\%. On SWE-bench, about 25\% of turns require
a filesystem checkpoint, while almost none require a full checkpoint.

\begin{figure}[tb]
  \centering
  \includegraphics[width=\linewidth]{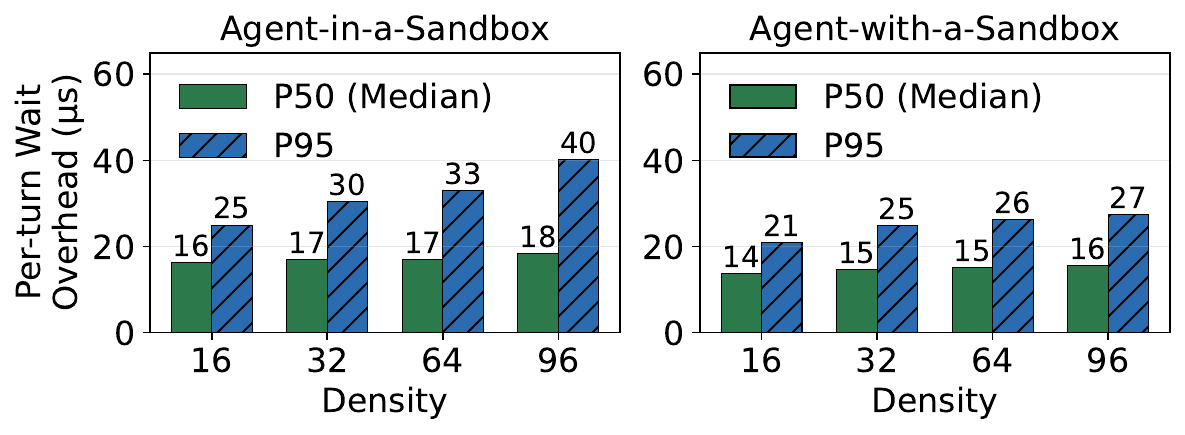}
  \vspace{-7mm}
  \caption{Per-turn Coordinator overhead. Even at 96-way co-location, the Coordinator adds only tens of \textit{microseconds} per turn, 4--5 orders of magnitude smaller than turn latency.}
  \Description{Per-turn coordinator overhead.}
  \label{fig:coordinator_bar}
\end{figure}

\subsection{Component Overhead}
\label{subsec:eval-overhead}

We break down the per-turn overhead of each component on \SystemName{}'s
critical path: the Coordinator, the Inspector, and checkpoint execution.

\PHM{Coordinator Overhead.} The Coordinator lies on the agent--LLM critical path
and must therefore be fast. Figure~\ref{fig:coordinator_bar} shows that its
per-turn overhead remains in the tens of \textit{microseconds} even at a
deployment density of 96 sandboxes per host: the median overhead is
18/16\,\textmu s for agent-in/with-a-sandbox deployments, and the p95 is
40/27\,\textmu s. Compared with LLM and tool latency
(Figure~\ref{fig:llm_tool_distribution_cdf}), this is 4--5 orders of magnitude
smaller and contributes less than 0.02\% overhead. The Coordinator is fast because it only forwards conversation metadata and
performs no data-intensive operations.

\begin{figure}[tb]
\centering
\includegraphics[width=\linewidth]{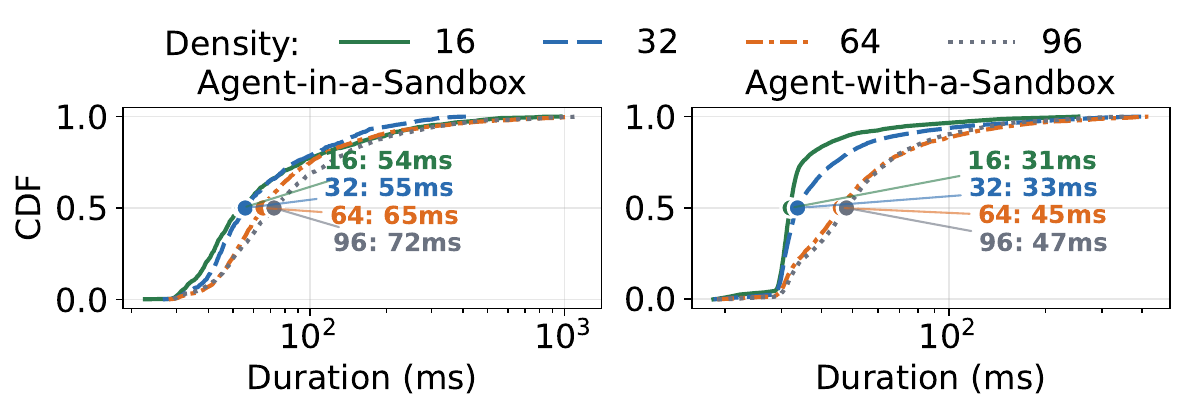}
\vspace{-8mm}
\caption{CDF of the Inspector's per-turn overhead across deployment densities 
for agent-in/with-a-sandbox.}
\Description{CDF of the Inspector's per-turn overhead across deployment densities 
for agent-in/with-a-sandbox.}
\label{fig:inspector_overhead_cdf}
\end{figure}

\PHM{Inspector Latency.} The Inspector runs \emph{asynchronously} after each
turn and is typically \textit{hidden behind the LLM wait window}, so its latency
is usually not exposed on the agent's critical path (\S\ref{subsec:inspector}).
Even in absolute terms, it remains modest: as shown in
Figure~\ref{fig:inspector_overhead_cdf}, for agent-in-a-sandbox deployment (left),
the median inspection latency is 54--72\,ms from 16 to 96 sandboxes per host;
for agent-with-a-sandbox deployment (right), it is 31--47\,ms. The p95 latency
stays below 200\,ms in all settings. Agent-in-a-sandbox latency is higher
because the eBPF hooks must also trace the agent process's syscalls inside the
sandbox, increasing event volume. This low overhead comes from a lightweight
design: eBPF records only a small amount of per-event state, and the daemon
performs simple per-sandbox net-change analysis in C.

\begin{figure}[tb]
\centering
\includegraphics[width=\linewidth]{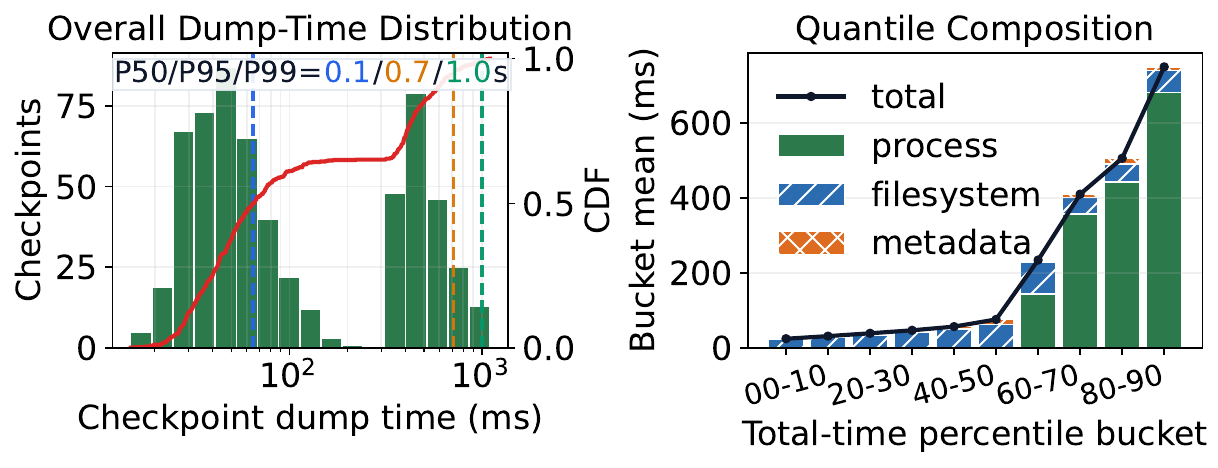}
\vspace{-8mm}
\caption{Checkpoint latency breakdown. \textbf{Left}: distribution of checkpoint latency. \textbf{Right}: checkpoint composition by latency quantile. Most checkpoints are fast filesystem-only dumps, while the long tail is dominated by process dumps.}
\vspace{-5mm}
\Description{Checkpoint time breakdown}
\label{fig:checkpoint_breakdown}
\end{figure}

\PHM{Checkpoint Latency Breakdown.} Checkpoint execution is also \emph{asynchronous}
and is often hidden within the LLM wait window, but its raw latency still
determines when overlap fails and delay becomes exposed.
Figure~\ref{fig:checkpoint_breakdown} (left) shows a clear bimodal distribution: one
mode at roughly 20--100\,ms and the other at 700--1000\,ms. The fast mode
corresponds to filesystem-only checkpoints, while the slow mode corresponds to
process checkpoints. Overall checkpoint latency has p50/p95/p99 of
0.1/0.7/1.0\,s. The composition plot (right) shows that up to the 70th percentile,
nearly all checkpoints complete within 50\,ms and are filesystem-only; the
near-second tail is almost entirely due to process checkpointing.

\subsection{Mechanism Effectiveness}
\label{subsec:eval-async}

We next evaluate how effectively \SystemName{} hides checkpoint latency behind
LLM waiting (\S\ref{subsec:coordinator}), and how much the C/R Engine's reactive scheduler
(\S\ref{subsec:cr-engine}) helps when that overlap becomes scarce. We use
Terminal-Bench for this study because it contains frequent process checkpoints
and therefore stresses the system more than SWE-bench, whose checkpoints are
almost exclusively filesystem-only.

\begin{figure}[tb]
  \centering
  \includegraphics[width=\linewidth]{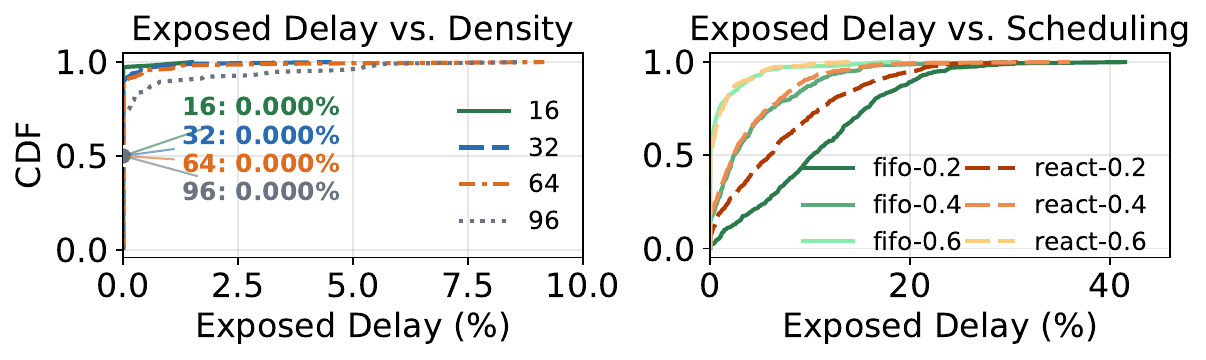}
  \vspace{-7mm}
  \caption{Effectiveness of asynchronous checkpointing and reactive scheduling. \textbf{Left}: task-level exposed delay under different sandbox densities. \textbf{Right}: task-level exposed delay under reduced LLM wait windows and different scheduling policies. The stress test scales the original LLM wait window to 0.2$\times$/0.4$\times$/0.6$\times$ of its original distribution at density 96.}
  \Description{Effectiveness of asynchronous checkpointing and reactive scheduling.}
  \label{fig:async_schedule_cdf}
\end{figure}

\PHM{Async Checkpointing.} We first measure, for each task, the fraction of
end-to-end runtime spent waiting at the completion gate for an unfinished
checkpoint. Figure~\ref{fig:async_schedule_cdf} (left) shows the CDF from 16 to
96 co-located sandboxes per host. Asynchronous checkpointing
(\S\ref{subsec:coordinator}) hides nearly all checkpoint cost: the median
exposed delay is zero across all densities, meaning that for at least half of
all tasks the checkpoint completes entirely within the LLM wait window. Even at
the 95th percentile, exposed delay remains small: 0.00\%, 0.37\%, 0.44\%, and
3.65\% at densities 16, 32, 64, and 96.

\PHM{Reactive Scheduling.} We next stress the scheduler under the hardest
setting: 96 sandboxes per host, with the LLM wait window scaled down to
0.2$\times$, 0.4$\times$, and 0.6$\times$ of its original distribution.
Smaller scaling factors both shorten the time available to overlap checkpointing
and increase checkpoint arrival rate by reducing turn duration.
Figure~\ref{fig:async_schedule_cdf} (right) compares the C/R Engine's reactive
scheduler with a FIFO baseline across all three scaling factors. Overall,
reactive scheduling reduces median exposed delay by up to 41.6\% and p95
delay by up to 31.3\%. The benefit comes from prioritizing checkpoints whose LLM
response has already returned, so work that has become exposed on the critical
path is completed before checkpoints that can still remain hidden.

\subsection{Case Study}
\label{subsec:eval-case-study}

\begin{figure}[tb]
  \centering
  \includegraphics[width=\linewidth]{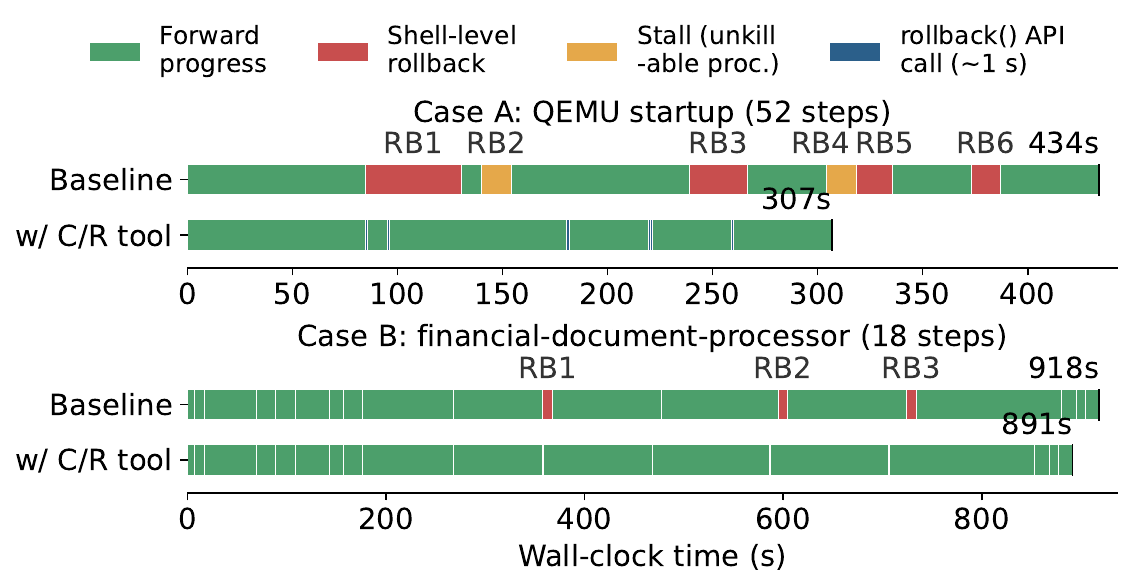}
  \vspace{-7mm}
  \caption{Proactive-rollback case studies: baseline vs.\ an agent equipped with a \texttt{rollback()} tool. \textbf{A}: QEMU startup, 434\,s\,$\rightarrow$\,307\,s ($-29\%$); RB2/RB4 stalls (orange) can be avoided. \textbf{B}: financial document classification, $-3\%$ wall clock but $-36\%$ rollback tokens.}
  \Description{Swim-lane comparison of two trajectories.}
  \vspace{-0.6cm}
  \label{fig:proactive_rollback}
\end{figure}

\PHB{Proactive Rollback.}
Conventional systems treat C/R as an agent-transparent safeguard. We ask whether exposing the C/R mechanism to the agent as a tool (\texttt{sbx.rollback(ckpt)}) can also \emph{enhance agentic execution} by replacing expensive shell-level self-recovery with a single O(1) API call when the agent corrupts its workspace. To study this question, we analyze two successful Terminal-Bench runs (Claude Opus-4.6) that nonetheless spend a large fraction of their step budget undoing earlier mistakes. Figure~\ref{fig:proactive_rollback} presents each run as a swim-lane. The \textit{Baseline} lane shows the executed trajectory. The \textit{w/~C/R~tool} lane replaces each detected rollback sequence with a single \texttt{rollback()} event at the measured p99 latency (1.00\,s), while keeping all forward-progress steps unchanged.

In \textit{Case A (QEMU startup)}, the agent boots an Alpine ISO using QEMU until \texttt{telnet~127.0.0.1~6665} returns a login prompt. The baseline completes in 52 steps (434\,s). Six rollback sequences consume 17 steps, 30.7\% of wall-clock time, and 50\% of the trajectory's tokens (14.3\,K of 28.7\,K). The main inefficiency is a \textasciitilde{}3-minute partial-cleanup (self-recovery) and stall in which an unkillable QEMU process keeps port 6665 occupied despite repeated \texttt{kill~-9} attempts (RB2/RB4). Replacing these sequences with \texttt{rollback()} shortens the trajectory by 29\%. More importantly, sandbox-level rollback restores a consistent known-good state and avoids such partial-cleanup stalls entirely.

In \textit{Case B (financial-document-processor)}, the agent classifies 17 mixed JPG/PDF documents. It runs four successive classifier scripts, each writing results directly into the target directories, while the first three are wrong. As a result, three rollback sequences (RB1--RB3) repeat the same cleanup boilerplate, \texttt{mv~FILES~/app/documents/;~rm~-f summary.csv}, consuming 22.8\,K of 62.9\,K incremental tokens (36\%). The wall-clock savings are modest (2.9\%) because these steps are filesystem-only; runtime is dominated by reasoning about the earlier error rather than by cleanup itself. However, the shell-level cleanup is brittle and sensitive to path changes, whereas sandbox-level rollback avoids this failure risk.

\begin{figure}[tb]
  \centering
  \includegraphics[width=\linewidth]{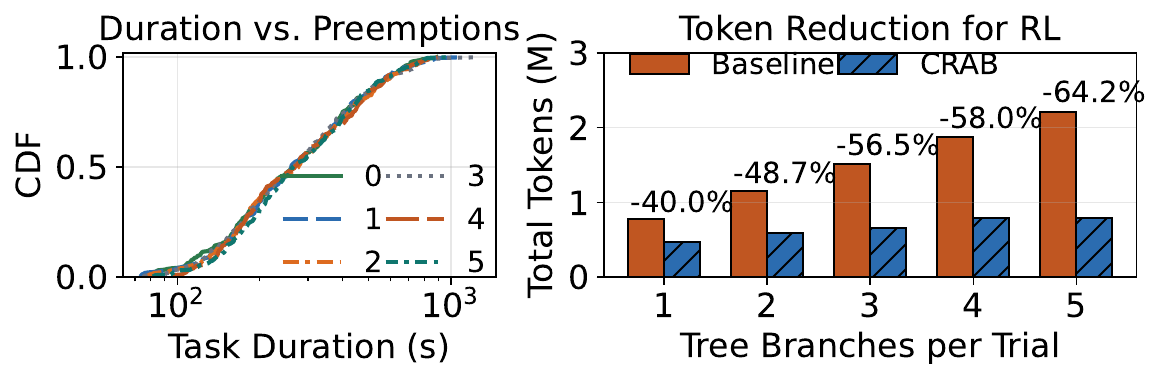}
  \vspace{-7mm}
  \caption{Case studies. \textbf{Left}: Task duration CDF at 1-5 preemption events injected to each task. \textbf{Right}: Token reduction for tree-based RL rollouts with 1-5 tree branches per trial.}
  \Description{Case study.}
  \label{fig:case_study}
\end{figure}

\PHM{Speculative Execution.} Another agent-facing use of C/R is \emph{speculative execution}. At each turn, a small and fast draft model proposes a candidate action, which \SystemName{} executes on a forked sandbox while the oracle model computes the ground-truth action in parallel. If the two actions match, \SystemName{} commits the fork's post-action state and skips re-executing the oracle action, thereby hiding the action latency behind the oracle model's inference time. If they differ, \SystemName{} discards the fork and executes the oracle action on the main sandbox. This use case requires efficient sandbox fork and rollback.

\begin{figure}[tb]
  \centering
  \includegraphics[width=0.9\linewidth]{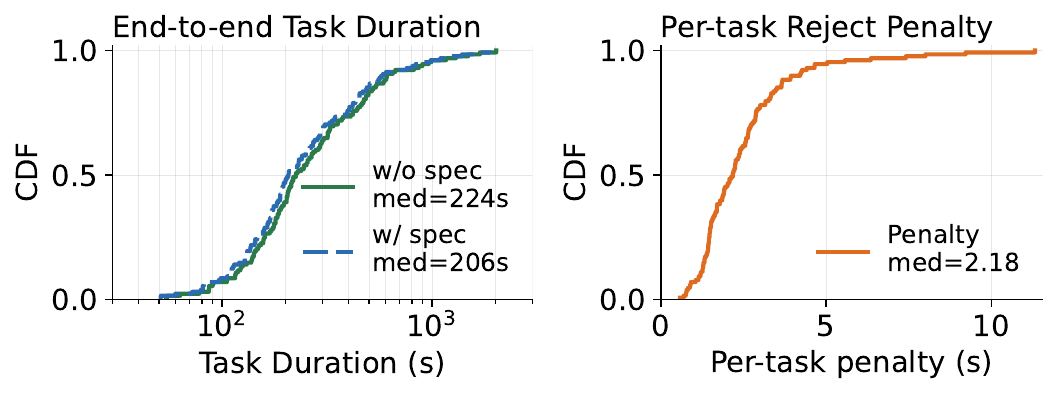}
  \vspace{-3mm}
  \caption{Speculative execution on SWE-Bench tasks.
  \textbf{Left:} Per-task wall-clock w/ and w/o speculation (observed). \textbf{Right:} Per-task penalty CDF (i.e., extra time the agent stalls due to rejected drafts.)}
  \Description{Two CDFs: end-to-end duration and loop-visible penalty.}
  \label{fig:speculative_exec}
\end{figure}

We evaluate this design on SWE-Bench tasks driven by SWE-agent. We emulate speculation with a draft model that is 10$\times$ faster than the oracle but achieves only about 50\% acceptance. As shown in Figure~\ref{fig:speculative_exec}, speculation reduces median task time from 224.1\,s to 206.5\,s, a 17.6\,s (7.9\%) improvement. The median penalty from mis-speculation is only 2.2\,s, or 0.9\% of total task time. Since the majority of turns are stateless, 58.0\% of fork requests during execution reuse the fork from the previous turn because the sandbox state has not changed, which avoids a fresh clone and restore.

\PHM{Spot Execution.} We next evaluate whether \SystemName{} can support
migration for preemptible spot-agent execution. In this setting, each sandbox
may be preempted with a 60\,s advance notice, with preemption events injected at
random. Upon receiving a preemption notice, \SystemName{} checkpoints the
sandbox on the preempted host; once a replacement instance is provisioned, the
checkpoint is restored on it. Checkpoint and restore together complete in under
1\,s (median/p95: 0.71/1.00\,s), so migration overhead is fully hidden whenever
provisioning finishes within the 60\,s grace period. To stress this path, we place
sandbox storage on a shared EBS volume with only 500\,MB/s peak bandwidth
(substantially slower than local NVMe SSDs), fix density at 96 sandboxes per
host, and vary preemption frequency so that each task experiences 1--5
preemptions. Figure~\ref{fig:case_study} (left) shows that, relative to a
no-preemption baseline, \SystemName{} adds only 0.45--3.01\% median and
1.01--7.30\% p95 time-to-solve, confirming that fast C/R supports spot execution
even under repeated preemptions and constrained storage bandwidth.

\PHM{RL Rollouts.} Finally, we evaluate whether \SystemName{} can reduce the
cost of tree-based RL post-training by reusing intermediate sandbox state across
branched rollouts~\cite{ji2025tree}. In this setting, the agent first explores
one rollout branch, then forks additional rollouts from a randomly chosen
intermediate point along the explored trajectory. We set the original batch size
to 16, so a branching factor of 4 yields an effective batch size of 64. Without
checkpoint reuse, each new rollout must re-execute the shared prefix; with
\SystemName{}, it resumes directly from the saved intermediate state.
Figure~\ref{fig:case_study} (right) varies the number of branches per trial from
1 to 5. Across them, \SystemName{} reduces rollout tokens by 40.0--64.2\%,
showing that efficient C/R can eliminate redundant prefix re-execution and
reduce RL post-training cost.




%% file: contents/7_RelatedWorks.tex
\section{Related Work}
\label{sec:related-work}

\PHB{Sandboxed Environments for Agents.} Agent platforms execute tools inside
isolated containers or
microVMs~\cite{e2b_docs,trenv-sosp24,openhands_docker_sandbox,yang2024swe},
focusing on provisioning, isolation, and reuse. \SystemName{} addresses an
orthogonal concern: \emph{when and what} to checkpoint by inferring
recovery-relevant state from OS-visible effects at turn boundaries.

\PHM{C/R and Persistent-State Recovery.} Checkpoint/restore has long served
migration, preemption, and failure recovery at the
process~\cite{criu_main,ansel2009dmtcp,hargrove2006berkeley},
container~\cite{runc_criu}, VM~\cite{firecracker-snapshot}, and whole-system
levels~\cite{treesls-sosp23}. Recent work targets modern workloads such as
differential checkpointing for distributed
training~\cite{checkfreq-fast21,checknrun-nsdi22,gemini-sosp23} and resilient LLM serving~\cite{dejavu,spotserve,failsafe,tarragon,faillite,anchortp}. These systems
checkpoint either full system state or application-managed state (e.g., DNN
weights). \SystemName{} instead \emph{infers} checkpoint granularity from
OS-visible turn effects---a mechanism absent from prior C/R work.

\PHM{Reducing Checkpoint and Snapshot Overhead.} Incremental
checkpointing~\cite{scr, criu_main}, asynchronous
flushing~\cite{fti-sc11}, and storage-aware
checkpointing~\cite{veloc-arxiv21,zfs_snapshots,btrfsSend} reduce per-snapshot
cost. These techniques are also applied in cloud serverless
systems~\cite{seuss-eurosys20,catalyzer-asplos20,faasnap-eurosys22,awsSnapStart}
to accelerate cold starts and migration. \SystemName{} is complementary: rather
than optimizing a single snapshot, it maintains evolving recovery points,
overlaps C/R with \emph{LLM wait windows}, and schedules checkpoint traffic at
host scope. Its eBPF-based Inspector draws on kernel-level tracing
techniques~\cite{linux_ebpf_syscall} but repurposes them to infer
recovery-relevant state at agent-turn granularity.

%% file: contents/8_Conclusion.tex
\section{Conclusion}
\label{sec:conclusion}

Agent sandboxes accumulate OS state that existing C/R mechanisms either fail to
capture or capture too aggressively. The root cause is an agent--OS semantic gap
that leaves neither layer with the information needed for efficient recovery.
\SystemName{} bridges this gap by observing that most agent turns produce no
recovery-relevant state, enabling a host-side runtime that infers checkpoint
granularity from OS-visible effects, overlaps C/R with LLM wait time, and
schedules checkpoint traffic across co-located sandboxes. On shell-intensive and
code-repair workloads, \SystemName{} raises recovery correctness from 8\% to
100\%, cuts checkpoint traffic by up to 87\%, and stays within 1.9\% of
fault-free execution time. More broadly, our results suggest that OS-visible
effects of opaque workloads can serve as lightweight semantic signals for
adaptive state management---a principle applicable beyond C/R to scheduling,
resource isolation, and security auditing of agent sandboxes.